\PassOptionsToPackage{dvipsnames}{xcolor}
\documentclass[twocolumn, linenumbers]{aastex631}
\watermark{}

\usepackage{acronym}

\usepackage{xspace}
\usepackage[normalem]{ulem}
\usepackage{amsmath}
\usepackage{ragged2e}

\newacro{imbh}[IMBH]{intermediate-mass black hole}
\newacro{bhns}[BHNS]{black hole neutron star}
\newacro{bbh}[BBH]{binary black hole}
\newacro{bh}[BH]{black hole}
\newacro{bns}[BNS]{binary neutron star}
\acrodef{FAR}[FAR]{false alarm rate}
\newacro{bf}[BF]{Bayes' factor}
\newacro{cbc}[CBC]{compact binary coalescence}
\newacro{ce}[CE]{Cosmic Explorer}
\acrodef{SNe}[SNe]{Supernova}
\newacro{da}[DA]{data analysis}
\newacro{et}[ET]{Einstein Telescope}
\newacro{eob}[EOB]{Effective-One-Body}
\newacro{fd}[FD]{frequency domain}
\newacro{gw}[GW]{gravitational-wave}
\newacro{gr}[GR]{general relativity}
\newacro{hm}[HM]{Higher mode}
\newacro{ifo}[IFO]{interferometer}
\newacro{imr}[IMR]{inspiral-merger-ringdown}
\newacro{im}[IM]{inspiral-to-merger}
\newacro{kagra}[KAGRA]{Kamioka Gravitational Wave Detector}
\newacro{ligo}[LIGO]{Laser Interferometer Gravitational-Wave Observatory}
\newacro{lso}[LSO]{Last Stable Orbit}
\newacro{lvc}[LVC]{LIGO-Virgo Collaboration}
\newacro{lvk}[LVK]{LIGO-Virgo-Kagra Collaboration}
\newacro{lo}[LO]{leading order}
\newacro{ns}[NS]{neutron star}
\newacro{nr}[NR]{numerical relativity}
\newacro{pn}[PN]{post-Newtonian}
\newacro{pe}[PE]{parameter estimation}
\newacro{psd}[PSD]{power spectral density}
\newacro{asd}[ASD]{amplitude spectral density}
\acrodef{kn}[KN]{kilonova}
\newacro{xg}[XG]{next-generation}
\newacro{jsd}[JSD]{jensen shannon divergence}
\newacro{sgrb}[sGRB]{short gamma ray burst}
\newacro{igwn}[IGWN]{international gravitational wave network}

\newacro{qc}[QC]{quasi-circular}
\newacro{snr}[SNR]{signal-to-noise ratio}
\acrodef{SNR}[SNR]{signal-to-noise ratio}
\newacro{ng}[NG]{Next Generation}
\newacro{eos}[EoS]{equation of state}
\newacro{em}[EM]{electromagnetic}

\newcommand{\BILBY}{\textsc{\texttt{Bilby}}\xspace}

\begin{document}


\title{The Critical Role of LIGO-India in the Era of Next-Generation Observatories}

\author{Shiksha Pandey}
\email{spp5950@psu.edu; alternate: shiksha1pandey@gmail.com}
\affiliation{Institute for Gravitation \& the Cosmos, Department of Physics, Penn State University, University Park PA 16802, USA}

\author{Ish Gupta}
\affiliation{Institute for Gravitation \& the Cosmos, Department of Physics, Penn State University, University Park PA 16802, USA}

\author{Koustav Chandra}
\affiliation{Institute for Gravitation \& the Cosmos, Department of Physics, Penn State University, University Park PA 16802, USA}

\author{Bangalore S. Sathyaprakash}
\affiliation{Institute for Gravitation \& the Cosmos, Department of Physics, Penn State University, University Park PA 16802, USA}
\affiliation{Department of Astronomy and Astrophysics, Penn State University, University Park PA 16802, USA}
\date{\today}

\begin{abstract}

We examine the role of LIGO-India in facilitating multi-messenger astronomy in the era of next-generation observatories. A network with two L-shaped Cosmic Explorer (CE) detectors and one triangular Einstein Telescope (ET) would precisely localize nearly the entire annual binary neutron star merger population up to a redshift of 0.5—over 10,000 events would be localized within \(10\,\mathrm{deg}^2\), including approximately 150 events within \(0.1\,\mathrm{deg}^2\). Luminosity distance would be measured to within 10\% for over 9,000 events and within 1\% for $\sim 100$ events. Surprisingly, replacing the 20 km CE detector with LIGO-India operating at A$^\sharp$ sensitivity (I$^\sharp$) yields nearly identical performance. The factor-of-five shorter arms are offset by a fourfold increase in baseline relative to a second CE in the U.S., preserving localization accuracy, with over 9,000 events within \(10\,\mathrm{deg}^2\) and $\sim 90$ events within \(0.1\,\mathrm{deg}^2\). This configuration detects $\sim 6,000$ events with luminosity distance uncertainties under 10\%, including $\sim 50$ with under 1\%. Both networks provide $\mathcal{O}(100)$ early-warning detections up to 10 minutes before merger, with localization areas $\leq 10 \deg^2$. While I$^\sharp$ enables excellent localization and early warnings, its shorter arms and narrower sensitivity band would limit its reach for other science goals, such as detecting population III binary black hole mergers at \(z \gtrsim 10\), neutron star mergers at \(z \sim 2\), or constraining cosmological parameters.
\end{abstract}

\section{Introduction}

\Acfp{gw} from inspiralling compact binaries offer a direct means to study compact objects like \acp{bh} and \acp{ns}. 
Predicted in 1915 by Einstein’s theory of \ac{gr}, \acp{gw} were first directly detected in 2015 by the Advanced LIGO detectors, marking the dawn of \ac{gw} astronomy~\citep{Aasi_2015, LIGOScientific:2016aoc}. In 2017, the Advanced Virgo detector joined this \ac{igwn}~\citep{Acernese_2015}. Their first joint detection, GW170814—a \ac{bbh} merger, was localised to a 90\% credible sky area of $\Delta \Omega_{90}=60\deg^2$, a significant improvement over the $1160\deg^2$ localization without Virgo~\citep{Abbott2017}.

Notably, the precise localization of GW170817, the first \ac{bns} merger event, was made possible because it was detected when both the Advanced LIGO detectors were observing~\citep{Abbott_2017}. This allowed for a $\Delta \Omega_{90} = 33 \deg^2$, enabling \ac{em} follow-up observations across the spectrum—from gamma rays to radio waves~\citep{Abbott_2017_mma, Goldstein_2017}. These \textit{multi-messenger} observations helped establish \ac{bns} mergers as major sites for r-process nucleosynthesis and progenitors of \ac{sgrb}.~\citep{Abbott_2017_sgrb, Abbott_2017_r, Kasen_2017, Goldstein_2017}.

In contrast, GW190425, the second BNS event, was detected solely by LIGO Livingston---LIGO-Hanford was offline and the signal in VIRGO was too weak to trigger detection. This resulted in a localization uncertainty of $\Delta \Omega_{90} \approx 8000 \deg^2$, nearly a quarter of the sky, making \ac{em} follow-up challenging~\citep{Abbott_2020}. Therefore, improved detector sensitivity, wider baselines, and strategic network configurations are crucial for enhancing future localization accuracy. Ongoing efforts to upgrade current detectors and build new generation of observatories such as the \acf{et} and \acf{ce} aim to address these limitations~\citep{Cahillane_2022, Acernese_2023, Punturo_2010_EinsteinTelescope, Evans_2021_CosmicExplorer, Iyer_2011}.

LIGO-India, the third LIGO interferometer currently being constructed in Aundha, India, is poised to contribute significantly to this effort. Expected to be operational in the 2030s with $A+$ or $A^{\sharp}$ sensitivity (see Fig. \ref{fig:sensitivity}), this detector is projected to improve the localization accuracy of sources further as it will provide a wider baseline~\citep{Saleem_2021, Shukla_2024, Gupta2024}. Previous studies have analyzed the performance of \ac{gw} detector networks, demonstrating their potential to advance multi-messenger astronomy, precision cosmology, and tests of \ac{gr} \citep{Evans_2021_CosmicExplorer, Sathyaprakash_2010, Schutz_2011, Klimenko_2011, Sathyaprakash:2012jk, Maggiore:2019uih, Adhikari_2019, Hall_2019, Borhanian:2022czq, Branchesi:2023mws, Harshank2023, Gupta2024, Chen_2024, Soni2024}. This paper evaluates the performance of ten \ac{gw} detector networks consisting of LIGO-Livingston, LIGO-Hanford, LIGO-India, \ac{ce} and \ac{et}, to assess LIGO-India's impact on multi-messenger astronomy. In Appendix \ref{app:aus}, we investigate additional network configurations that incorporate a southern‑hemisphere observatory in New South Wales, Australia.

\begin{table}[htb!]
  \centering
  \caption{Planned Configurations of Gravitational Wave Detector Networks}
  \vspace{-2mm}
  \renewcommand{\arraystretch}{1.04}
  \begin{tabular}{c c}
    \hline\hline
    Configuration & Detectors Included \\
    \hline 
    CE40LI$+$ & 40 km CE + LLO in A$^\sharp$ + LIO in A$+$ \\
    CE4020I$+$ & 40 km CE + 20 km CE + LIO in A$+$ \\
    CE40I$+$ET & 40 km CE + ET + LIO in A$+$ \\
    \hline
    HLI$^\sharp$ & LHO in A$^\sharp$ + LLO in A$^\sharp$ + LIO in A$^\sharp$ \\
    CE40LI$^\sharp$ & 40 km CE + LLO in A$^\sharp$ + LIO in A$^\sharp$ \\
    CE4020I$^\sharp$ & 40 km CE + 20 km CE + LIO in A$^\sharp$ \\
    CE40I$^\sharp$ET & 40 km CE + ET + LIO in A$^\sharp$ \\
    \hline
    CE4020L & 40 km CE + 20 km CE + LLO in A$^\sharp$ \\
    CE40LET & 40 km CE + LLO in A$^\sharp$ + ET \\
    CE4020ET & 40 km CE + 20 km CE + ET \\
    \hline
  \end{tabular}
  \begin{flushleft}
\justifying\textbf{Notes.} CE and ET are considered at their design sensitivities, LIGO-Livingston (LLO) and LIGO-Hanford (LHO) at A$^\sharp$, and LIGO-India (LIO) either at A$+$ or A$^\sharp$ sensitivity. The 20 km CE is located off the coast of Texas in all configurations except CE4020L, where it is in the Great Lakes.
  \end{flushleft}
  \label{tab:detector_configurations}
\end{table}

\section{Detector Networks}
A single \ac{gw} interferometer provides limited positional information due to its slowly varying antenna pattern, which favors two broad regions perpendicular to the plane of the detector's arms. In contrast, a network of non-collocated, non-coaligned detectors offers significantly better sky localization by enabling triangulation. This is achieved through the time differences, phase differences, and amplitude ratios of the \ac{gw} signal as it arrives at each site~\citep{Fairhurst:2010is}.

\begin{figure}[ht]
    \centering
    \includegraphics[width=\columnwidth]{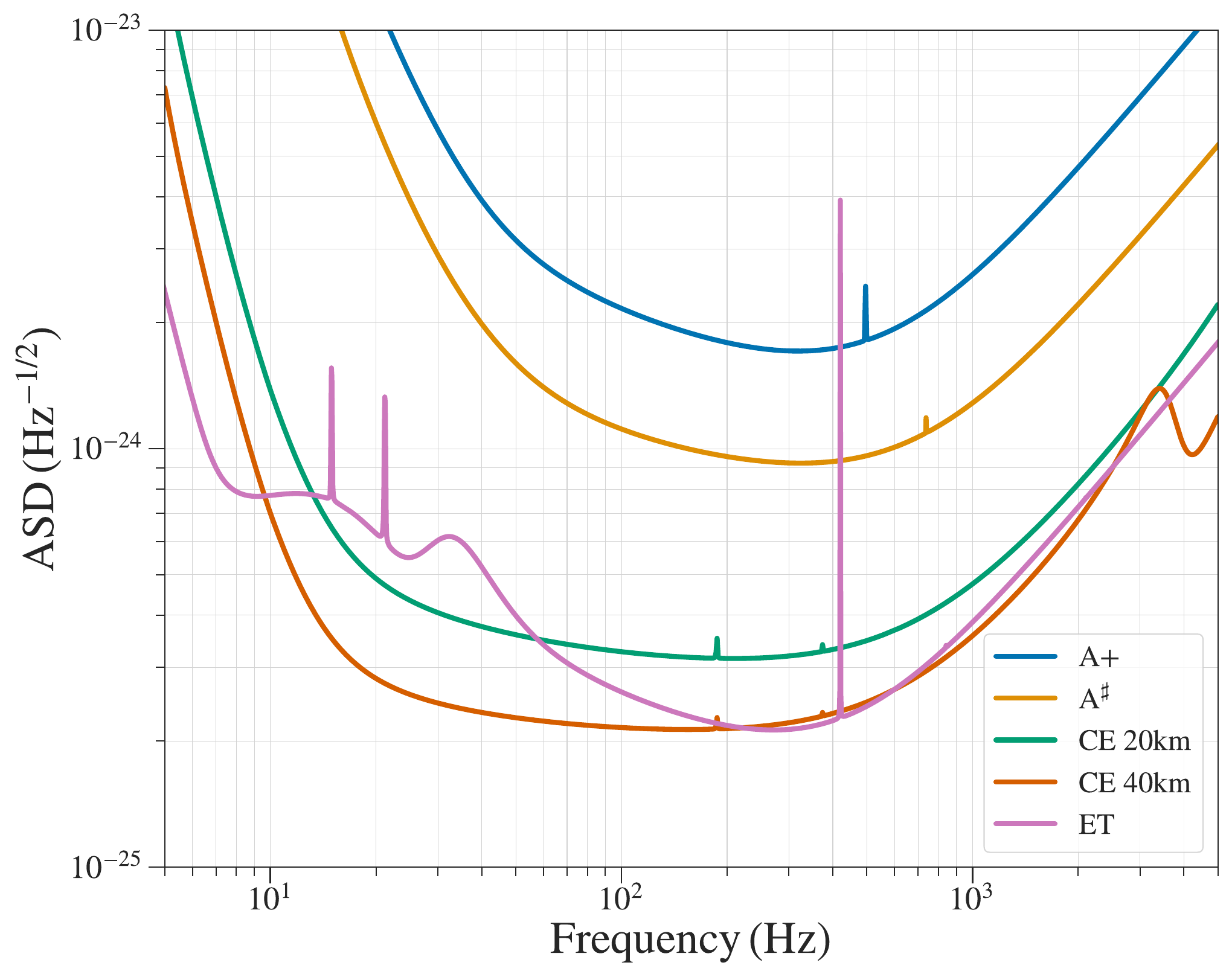}
    \caption{Amplitude spectral density (ASD) of detector noise for LIGO at A$+$ and A$^\sharp$ sensitivity, and 20km and 40km \ac{ce}, along with \ac{et}.}
    \label{fig:sensitivity}
\end{figure}

The network configurations considered in this study, incorporating both current and future \ac{gw} observatories, are summarized in Table \ref{tab:detector_configurations} and their design sensitivity curves are shown in Fig. \ref{fig:sensitivity}~\citep{Hild:2010id, T1800042-v5, LIGO-T1500293-v13, Evans_2021_CosmicExplorer, CE-T2000017, T2300041-v1, LIGO-T2200287}.

The LIGO observatory currently consists of two L-shaped interferometers located at Hanford and Livingston, each with 4 km arms. Operating in their Advanced configuration, these detectors can achieve strain sensitivities on the order of a few $10^{-24} / \sqrt{\text{Hz}}$. The A$+$ upgrade, scheduled for the fifth observing run, is expected to improve sensitivity by a factor of two, while the subsequent A$^\sharp$ upgrade aims to enhance sensitivity further, reaching $\sim 9 \times 10^{-25} / \sqrt{\text{Hz}}$ around 300 Hz. The third LIGO interferometer, currently under construction in India, will feature 4 km-long arms like its US counterparts and is expected to operate at A$+$ (and with upgrades at A$^\sharp$) sensitivity levels. Its geographical location will provide a wider baseline, significantly enhancing the global network’s triangulation capabilities.

The L-shaped \ac{ce}, with its proposed 40 km and/or 20 km arms, is expected to be built in the continental US~\citep{Reitze:2019iox, Evans_2021_CosmicExplorer, Evans:2023euw}. The triangular-\ac{et} with its 10km arms, on the other hand, is expected to be constructed in Europe; however, proposals exist for the construction of two L-shaped \acp{et}~\citep{Punturo_2010_EinsteinTelescope, Branchesi:2023mws}. Our study considers a triangular \ac{et} at Sardinia, Italy. 

These \ac{xg} detectors are expected to be at least ten times more sensitive than current detectors and will have larger frequency bandwidth. In particular, the \ac{asd} of \ac{ce} with 40 km arms, will be designed to reach values of $\sim 7 \times 10^{-25} / \sqrt{\text{Hz}}$ at 10 Hz and $\sim 2-3 \times 10^{-25} / \sqrt{\text{Hz}}$ over $20-800 \text{ Hz}$. \ac{et} is anticipated to outperform \ac{ce} at frequencies below 8 Hz. Together, they will increase the \ac{gw} detection rate by several orders of magnitude, with observable signals persisting for durations ranging from minutes to hours. 

\section{Inference of binary neutron star parameters}

We perform our analysis using \texttt{\textsc{GWBENCH}}, a Fisher information package for \ac{gw} benchmarking~\citep{Borhanian:2020ypi}, for a population of \ac{bns} mergers within $z=0.5$. Fisher matrix packages similar to \texttt{\textsc{GWBENCH}} have also been used in comparative studies of different detector networks ~\citep{Iacovelli:2022bbs, Dupletsa:2022scg}. Fisher matrix is a quadratic approximation that is only reliable in the strong signal limit. To understand the veracity of these results, we performed a tall Bayesian analysis for $\sim 7\%$ of these signals (see Section \ref{sec:bilby}). \texttt{\textsc{GWBENCH}} approximates the parameter measurement uncertainties using the Covariance matrix, which is the reciprocal of the Fisher information matrix defined as, 
\begin{equation}
    \Gamma_{ab} \equiv 2 \int_{f_{\text{min}}}^{f_{\text{max}}} \frac{\tilde{h}_{a} \tilde{h}_{b}^* + \tilde{h}_{a}^* \tilde{h}_{b}}{S_n^A(f)} \, df.
    \label{eq:fisher_matrix}
\end{equation}
Here, $\tilde{h}_{a}$ is the partial derivatives of the \ac{gw} signal with respect to the parameter $\lambda_a,$ $\tilde h_a \equiv \partial h/\partial \lambda_a,$ $S_n^A(f)$ is the noise \ac{psd} of detector $A$ and $f_{\text{min}}$ and $f_{\text{max}}$ are the minimum and maximum frequency cutoffs, respectively, determined by a detector's sensitivity bandwidth. The diagonal elements of the covariance matrix describe the variance of signal parameters, while the off-diagonal elements describe the correlations between different parameters.

To ensure proper sampling of probability density functions (PDFs) in our analyses, we simulate $1.85\times 10^6$ BNS mergers with redshifts $z \leq 0.5$, distributed across five bins: [0.0005, 0.1], [0.1, 0.2], [0.2, 0.3], [0.3, 0.4], and [0.4, 0.5]. Although our simulated sample size is extensive, the redshift distribution is normalized to follow Eq. \ref{eq:md_rate}. Specifically, integrating Eq. \ref{eq:md_rate} over $z \leq 0.5$ predicts a total of 16,585 BNS mergers annually. The component masses, $m_k$, are drawn from a uniform distribution within the range $[1,\ 2.2]\,M_\odot$, and the spin magnitudes, $a_k$, are selected from within $[0,\ 0.1]$. The tidal deformability of the \ac{ns} is modeled using the APR4 \ac{eos}~\citep{Akmal_1998}, and the \ac{gw} signals are generated using the IMRPhenomPv2\_NRTidalv2 waveform approximant~\citep{Dietrich:2019kaq}. Using these simulations, we calculate detection rates, \ac{snr} ($\rho$), sky localization area ($\Delta \Omega_{90}$), parameter estimation accuracies, and early warning times.

\section{Detection rate}

The detection rate, defined as:
\begin{equation}
    D_R (z, \rho^*) = \int_0^z  \epsilon(z^{\prime}, \rho^*) \frac{\dot{n}(z')}{1 + z'} \frac{dV_c}{dz'} dz',
\end{equation}
measures the cumulative number of observable \ac{bns} mergers up to a redshift \(z\). Here, \(\epsilon(z, \rho^*)\) is the detection efficiency (which is the fraction of all gravitational wave sources within a redshift of $z$ detected with an \ac{snr} larger than the threshold $\rho^*$, see Appendix \ref{app:efficiency}), $\dot{n}(z)$ is the source-frame merger rate density at redshift $z$, the factor \((1 + z)^{-1}\) accounts for cosmological time dilation and \(dV_c/dz\) is the differential comoving volume element. The merger rate density $\dot{n}(z)$ is modeled using the Madau-Dickinson star formation rate $\psi(z)$ \citep{Madau_2014}:
\begin{align}
\psi(z) &= (1+z)^{\gamma}\left [1+\left(\frac{1+z}{1+z_p}\right)^{\kappa} \right ]^{-1}\\
\dot{n}(z) &= A \int_{t_d^{\text{min}}}^{t_d^{\text{max}}} \psi(z_f(z, t_d)) P(t_d) \, dt_d,
\label{eq:md_rate}
\end{align}
with \(z_p = 1.9\) (the redshift corresponding to the peak star formation rate), \(\gamma = 2.7\), and \(\kappa = 5.6\). The parameter $t_d$ denotes the time delay from the formation of a binary at redshift $z_f$ to its merger at $z$, $P(t_{d}) \propto 1/td$ is the delay time distribution (taken to be Jeffry's prior \citep{DE_FREITAS_PACHECO_2006}), and the integration limits, $t_d^{min} = 20 \text{ Myr}$ and $t_d^{max} = 10 \text{ Gyr}$, represent the minimum and maximum time delays. The normalization constant $A$ is chosen so that the local merger rate density $\dot{n}(0)= 320 \ \mathrm{Gpc}^{-3} \mathrm{yr}^{-1}$ \citep{Abbott_2021}. For further details on this derivation, see \citep{Regimbau_2009, Borhanian:2022czq}.

\begin{figure}[hbt]
  \centering
  \makebox[.475\textwidth]{%
        \includegraphics[width=.475\textwidth]{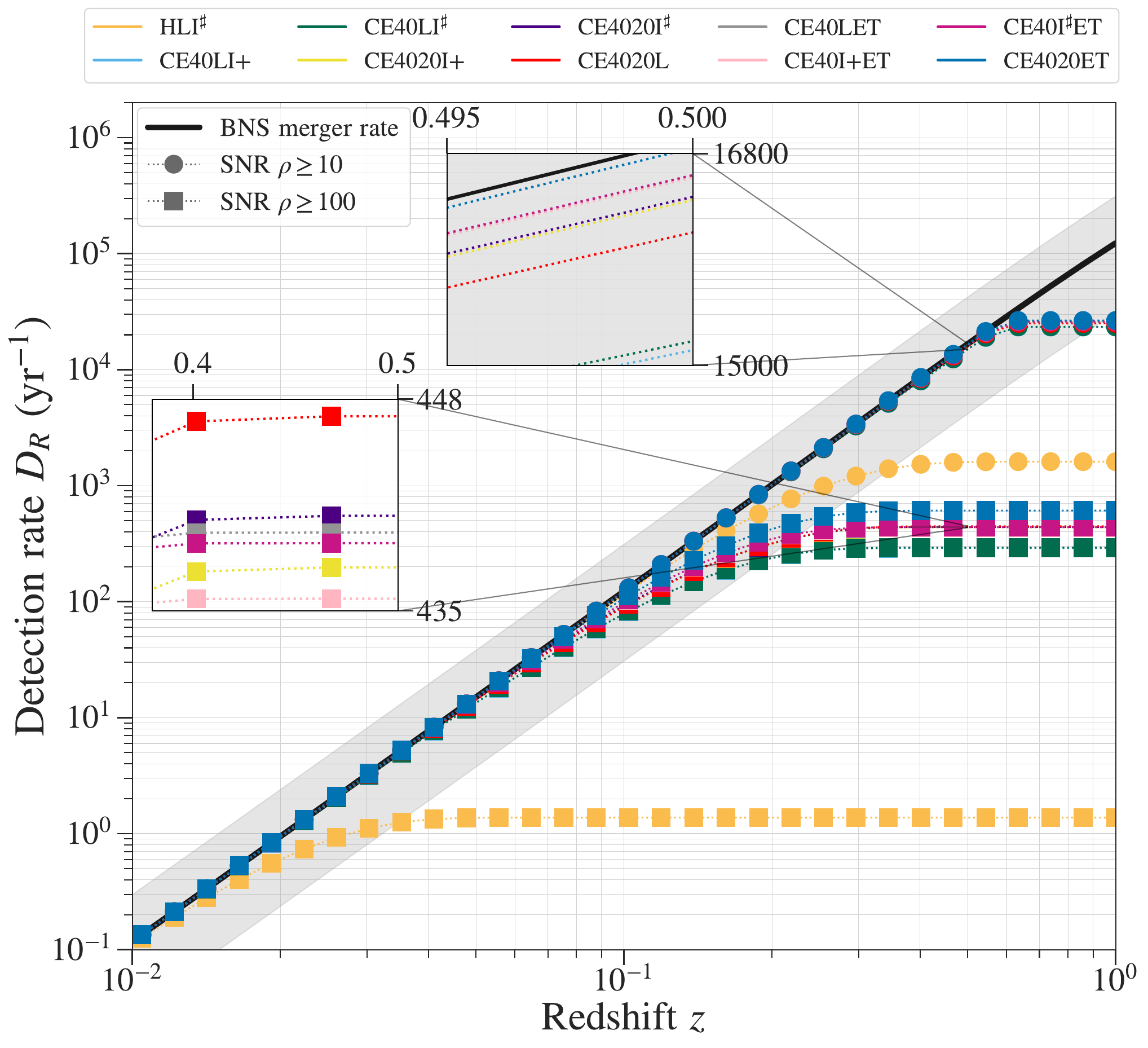}}
  \caption{Detection rate as a function of redshift for ten GW detector networks. Solid circles and squares represent detection rates evaluated with SNR thresholds of 10 and 100, respectively. The solid black line is the cosmic BNS merger rate, and the gray-shaded region is the variation in the merger rate due to current uncertainties in local merger rate density. For both SNR thresholds, CE4020ET exhibits the highest detection rate, while HLI$^\sharp$ shows the lowest. For $\ac{SNR} \geq 10$, nearly all networks (except HLI$^\sharp$) have similar rates and thus overlap in the main plot. For $\ac{SNR} \geq 100$, CE40LI+ and CE40LI$\sharp$ are indistinguishable, and the rate curves for CE4020L, CE4020I+, CE4020I$^\sharp$, CE40LET, and CE40I+ET are hidden behind that of CE40I$^\sharp$ET. Two zoom-in insets highlight these regions of overlap.
}
  \label{fig:detection}
\end{figure}

\begin{figure*}[htb!]
  \centering
  \includegraphics[width=0.78\textwidth]{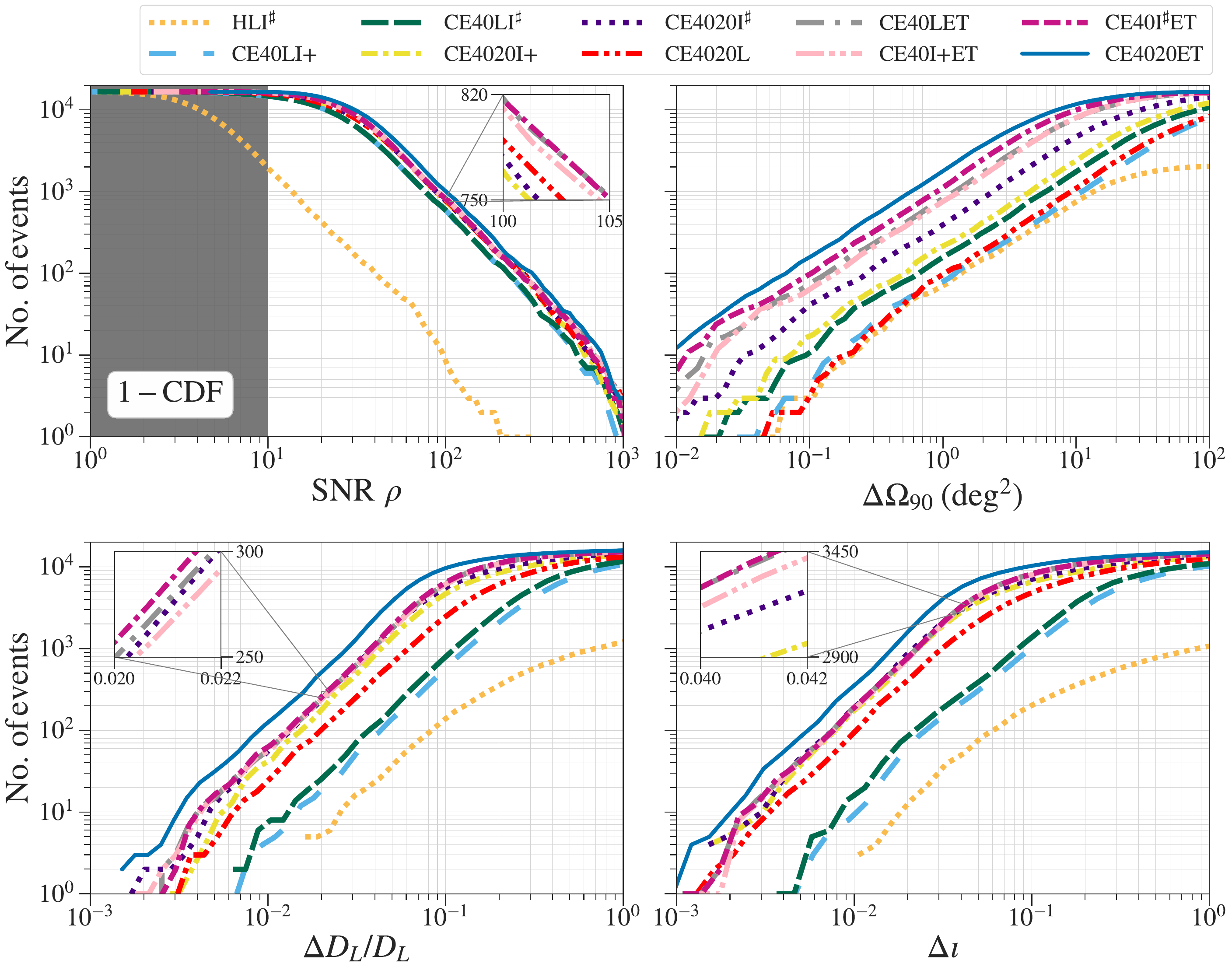}
  \caption{Comparative analysis of multi-messenger observational parameters for BNS mergers across ten gravitational wave detector network configurations. The figure shows the \ac{SNR} histogram (top left), 90\% credible sky localization area (top right), and uncertainties in the relative distance (bottom left) and the inclination angle (bottom right).
All uncertainties are presented at the 68\% credibility level. All panels, except for the \ac{SNR}, include events with SNR$\geq 10$, ensuring that localization and parameter estimation comparisons are made for confidently detected signals. Across all metrics, CE4020ET performs best, followed closely by CE40I$^\sharp$ET. CE40LI+ and CE40LI$^\sharp$ show nearly indistinguishable detection performance. The luminosity distance and inclination angle uncertainties for CE40I$^\sharp$ET, CE40I+ET, CE40LET, CE4020I+, and CE4020I$^\sharp$ are similar. Insets reveal network curves that are otherwise obscured due to overlapping traces.
}
  \label{fig:visible_mma}
\end{figure*}

Fig.~\ref{fig:detection} shows the expected annual \ac{bns} merger detection rate. The solid black line represents the cosmic \ac{bns} merger rate, which is a theoretical upper limit on the number of events that could be detected, given no limitations in detector sensitivity or range. The gray-shaded region indicates the uncertainty in the local merger rate $\dot n(0)$, reflecting variations in astrophysical estimates and models. Up to $z=0.1$, all configurations can detect more than 100 events annually with $\rho > 100,$ except for the HLI$^\sharp$ network which observes only two events. As $z$ increases, $D_R (z, \rho^*)$ increases for all networks due to the expanding comoving volume and the corresponding increase in merger events. Most networks closely approach or align with the cosmic BNS merger rate throughout the redshift range examined for \(\rho^* = 10\). HLI$^\sharp$, however, begins to plateau around $z=0.3$ at this \ac{snr} threshold, indicating its limited ability to observe distant events. For \(\rho^* = 100\), this saturation occurs even earlier at $z = 0.07$, while other configurations begin to diverge from the theoretical maximum after $z = 0.1$. These results indicate that while a traditional setup is proficient at detecting nearby \ac{bns} mergers, only those incorporating \ac{xg} detectors exhibit sustained effectiveness at greater distances, underscoring their importance in extending the detectable horizon.

\section {Precision in Parameter Estimation}
Figure \ref{fig:visible_mma} shows the number of events as a function of \ac{snr} $\rho$, sky localization area at 90\% credible interval $\Delta \Omega_{90}$, relative luminosity distance measurement uncertainty $\Delta D_L/D_L$, and inclination angle measurement uncertainty $\Delta \iota$. 

We see that at $\rho < 30$, most networks observe a large number of events, suggesting robust sensitivity to weaker signals. For higher \ac{snr} values, i.e., for high-fidelity sources, the performance of the HLI$^\sharp$ network declines considerably with no event detected at  \(\rho > 300\). The nearly overlapping curves for the rest of the network configurations across the given \ac{snr} range demonstrate their comparable sensitivity. The top right panel presents the performance of different networks in accurately determining the sky position. Only CE4020ET, CE40IET, CE40LET and CE4020I configurations can localize $> 10$ events within a 90\% credible interval of $\Delta \Omega_{90} < 0.1 \deg^2$. For a $\Delta \Omega_{90} < 1 \deg^2$, event counts range from $< 100$ for HLI$^\sharp$, CE40LI+, and CE4020L to $ \sim 1000$ for CE4020ET and CE40I$^\sharp$ET, with networks that incorporate at least one \ac{xg} detector alongside LIGO-India showing significantly improved performance. It is important to note that although both CE4020L and HLI$^\sharp$ demonstrate similar 2D localization, CE4020L distinctly surpasses HLI$^\sharp$ in accurately measuring $D_L$ and $\iota$, and the story is similar for CE40LI+. At $\Delta \Omega_{90} \approx 100 \deg^2$, nearly all configurations can adequately localize \ac{bns} mergers. Although such a broad localization area reduces the precision of follow-up observations, it is still beneficial for initial detections and sending alerts to the astronomical community.

\section{Observational Imperatives for EM Follow-ups }
\Acfp{sgrb} may occur within seconds to minutes following a \ac{bns} merger, likely driven by the formation of a relativistic jet that breaks through the surrounding dynamical ejecta~\citep{Rezzolla_2011, Berger_2014, Ghirlanda_2019}. These \acp{sgrb} are frequently accompanied by \ac{em} afterglows observable across multiple wavelengths, including X-ray, optical, and radio bands, resulting from the interaction between the jet and the ambient medium~\citep{Troja_2017, Hallinan_2017, Margutti_2017, Margutti_2018}.

In addition, \ac{kn}—transient optical and infrared emissions—arise from the radioactive decay of r-process elements synthesized in the neutron-rich ejecta. \Ac{kn} light curves typically peak in the optical band within 1-2 days and in the infrared band within 5-7 days post-merger~\citep{Pian_2017, Kasen_2017, Radice_2018, Metzger_2019}. These observations are essential for probing the nature of matter at supranuclear densities, where exotic states of matter may emerge~\citep{Abbott_2018, Dietrich_2020, Annala_2018, De_2018}.

\Ac{em} observations also provide valuable insights into the properties of host galaxies, providing clues about the formation channels and evolutionary pathways of \ac{bns} systems, their merger rates and their role in the chemical enrichment of the universe~\citep{Fong_2013, Berger_2014, Levan_2017, C_t__2019}. Successful observation and characterization of these phenomena rely on precise three-dimensional (3D) localization of \ac{bns} mergers and the deployment of efficient early warning systems, which enable prompt follow-up observations by electromagnetic telescopes ~\citep{Chan:2018csa, Nitz:2018rgo,Hanna:2019ezx, Sachdev:2020lfd, Li:2021mbo, Tohuvavohu:2024flg}. 
This section discusses different detector networks' 3D source localization and early warning capabilities.

\subsection{3D Source Localization}

\begin{table*}[t]
\centering
\caption{Annual BNS Merger Count and Redshift Reach for Considered Networks, Corresponding to Specific $\Delta\Omega_{90}$}
\vspace{-2mm}
\label{tab:no_sky_area}
\renewcommand{\arraystretch}{1.04}
\fontsize{9.6}{12}\selectfont
\begin{tabular}{ c  c  c  c  c  c  c  c  c  c  c }
\toprule
Quantity & \ 40LI$+$ \ & \ 4020I$+$ \ & \ 40I$+$ET \ & \ HLI$^\sharp$ \ & \ 40LI$^\sharp$ \ & \ 4020I$^\sharp$ \ & \ 40I$^\sharp$ET \ & \ 4020L \ & \ 40LET \ & \ 4020ET \ \\
\hline
\multicolumn{11}{c}{$\Delta\Omega \leq 1$ deg$^2$} \\
\hline
Number & 74 & 203 & 672 & 69 & 146 & 356 & 987 & 90 & 733 & 1578\\
Median $z$ & 0.032 & 0.056 & 0.094 & 0.031 & 0.047 & 0.073 & 0.115 & 0.041 & 0.099 & 0.148\\
Maximum $z$ & 0.091 & 0.136 & 0.263 & 0.091 & 0.118 & 0.203 & 0.347 & 0.444 & 0.299 & 0.496\\
\hline
\multicolumn{11}{c}{$\Delta\Omega \leq 10$ deg$^2$} \\
\hline
Number & 840 & 2214 & 7309 & 710 & 1606 & 4339 & 9277 & 1011 & 7236 & 11090\\
Median $z$ & 0.108 & 0.179 & 0.286 & 0.102 & 0.151 & 0.241 & 0.312 & 0.129 & 0.287 & 0.329 \\
Maximum $z$ & 0.299 & 0.496 & 0.5 & 0.41 & 0.438 & 0.499 & 0.5 & 0.466 & 0.5 & 0.5 \\
\hline
\multicolumn{11}{c}{$\Delta\Omega \leq 100$ deg$^2$} \\
\hline
Number & 7477 & 11809 & 15951 & 2028 & 10372 & 14087 & 16139 & 7941 & 15897 & 16454\\
Median $z$ & 0.298 & 0.359 & 0.367 & 0.161 & 0.327 & 0.355 & 0.368 & 0.309 & 0.366 & 0.37\\
Maximum $z$ & 0.5 & 0.5 & 0.5 & 0.466 & 0.5 & 0.5 & 0.5 & 0.5 & 0.5 & 0.5\\
\hline
\end{tabular}
\begin{flushleft}
\justifying\textbf{Notes.} Redshift is capped at $z = 0.5$. We choose a median local merger rate density of 320 Gpc$^{-3}$yr$^{-1}$ \citep{Abbott_2021}. "CE"is ommitted from network names for brevity (e.g., CE4020L is referred to as 4020L here).
\end{flushleft}
\end{table*}

\begin{figure*}[htb!]
  \centering
  \includegraphics[width=1.0\textwidth]{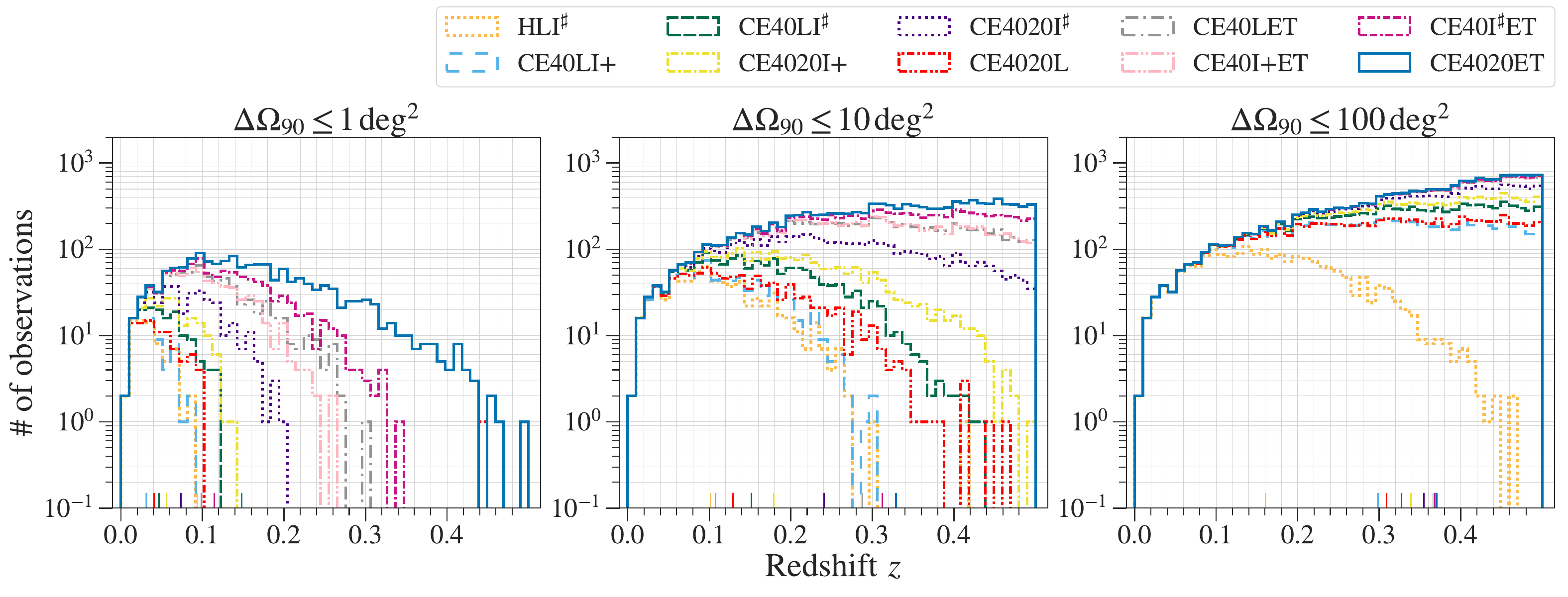}
\caption{The figure displays (from left to right) the number of events that can be localized to within sky areas of 1, 10, and 100 square degrees, respectively, as a function of redshift. In the  $1\ \mathrm{and}\ 10\ \mathrm{deg}^2$ panels, CE40LI+ and HLI$^\sharp$ show nearly coincident curves, whereas at $100\ \mathrm{deg}^2$ CE40LI+ significantly outperforms HLI$^\sharp$. Similarly, CE40I+ET and CE40LET localize a comparable number of events to within 10 square degrees, and CE4020ET, CE40I$^\sharp$ET, CE40I+ET, CE40LET show similar performance at 100 square degrees. Overall, CE4020ET precisely localizes the highest number of events followed by CE40I$^\sharp$ET. The variation in localization accuracy across networks, with some maintaining high precision at higher redshifts while others show a rapid decline, underscores the role of network design in counterpart identification (see text for details).}
  \label{fig:sa90}
\end{figure*}

Table \ref{tab:no_sky_area} and Figure~\ref{fig:sa90} show the number of detectable \ac{bns} mergers per year and the median and maximum redshift (up to $z=0.5$) for these events corresponding to $\Delta \Omega_{90} \leq 1,\ 10\ \mathrm{and}\ 100 \deg^2$, respectively. Figure \ref{fig:visible_mma} (third panel) provides information on uncertainties in luminosity distance ($D_L$) measurements. The CE4020ET configuration consistently detects the highest number of events for all considered sky localization areas, with greater median and maximum redshifts, indicating its ability to reach the farthest distances. CE40I$^\sharp$ET closely follows in performance. For instance, CE4020ET (CE40I$^\sharp$ET) localizes 11,090 (9277) events within $10 \deg^2$, and detects 9,019 (5790) events with $\Delta D_L/ D_L < 10\%$, including 110 (55) events with sub-1\% error in $D_L$. They also identify around 146 (86) events within $\Delta \Omega = 0.1 \deg^2$. CE40I+ET and CE40LET trail slightly behind CE40I$^\sharp$ET. 

In contrast, CE40LI networks detect an order of magnitude fewer events at similar luminosity distance error thresholds compared to CE40I$^\sharp$ET setup. Moreover, HLI$^\sharp$ and CE4020L networks detect significantly fewer events--710 (1011) that can be localized within $10 \deg^{2}$--with a median redshift of 0.102 for HLI$^\sharp$ and 0.129 for CE4020L. HLI$^\sharp$ detects only around 120 events with 10\% error in $D_L$ and none with sub-1\% precision. These networks localize only 3 events each within $0.1 \deg^2$. 

The superior 3D localization capabilities of networks such as CE4020ET, CE40IET, and CE40LET significantly improve the prospects for \ac{em} counterpart observations, allowing for more efficient coordinated searches with telescopes. This facilitates host galaxy identification and subsequently the measurement of the Hubble constant~\citep{Chen:2024gdn, Chandra:2024dhf}. For instance, with the Rubin Observatory’s FoV spanning $9.6 \deg^2$ \citep{Ivezi__2019}, it is possible to observe $\mathcal{O}(10^4)$ events yearly. While telescopes with narrower FoVs would require multiple pointings to cover the same sky area, they can still provide valuable contributions. However, \ac{gw} networks with fewer well-localized events present challenges for detecting \ac{em} counterparts, even for wide-field instruments such as Rubin. This challenge is compounded for smaller telescopes, increasing the observational burden and potentially delaying follow-up efforts.

\subsection{Early Warning Systems}
With alert latencies typically ranging from 30 seconds to a few minutes after GW detections, early warnings allow for the timely positioning of \ac{em} instruments, ensuring that initial moments of the merger and associated \ac{em} counterparts are observed \citep{Chan:2018csa, Nitz:2018rgo,Hanna:2019ezx, Sachdev:2020lfd, Li:2021mbo, Tohuvavohu:2024flg}. The rapid response is essential to capture transient phenomena like short gamma-ray bursts (\ac{sgrb}), which fade rapidly.   

\begin{figure*}[hbt!]
  \centering
  \includegraphics[width=0.78\textwidth]{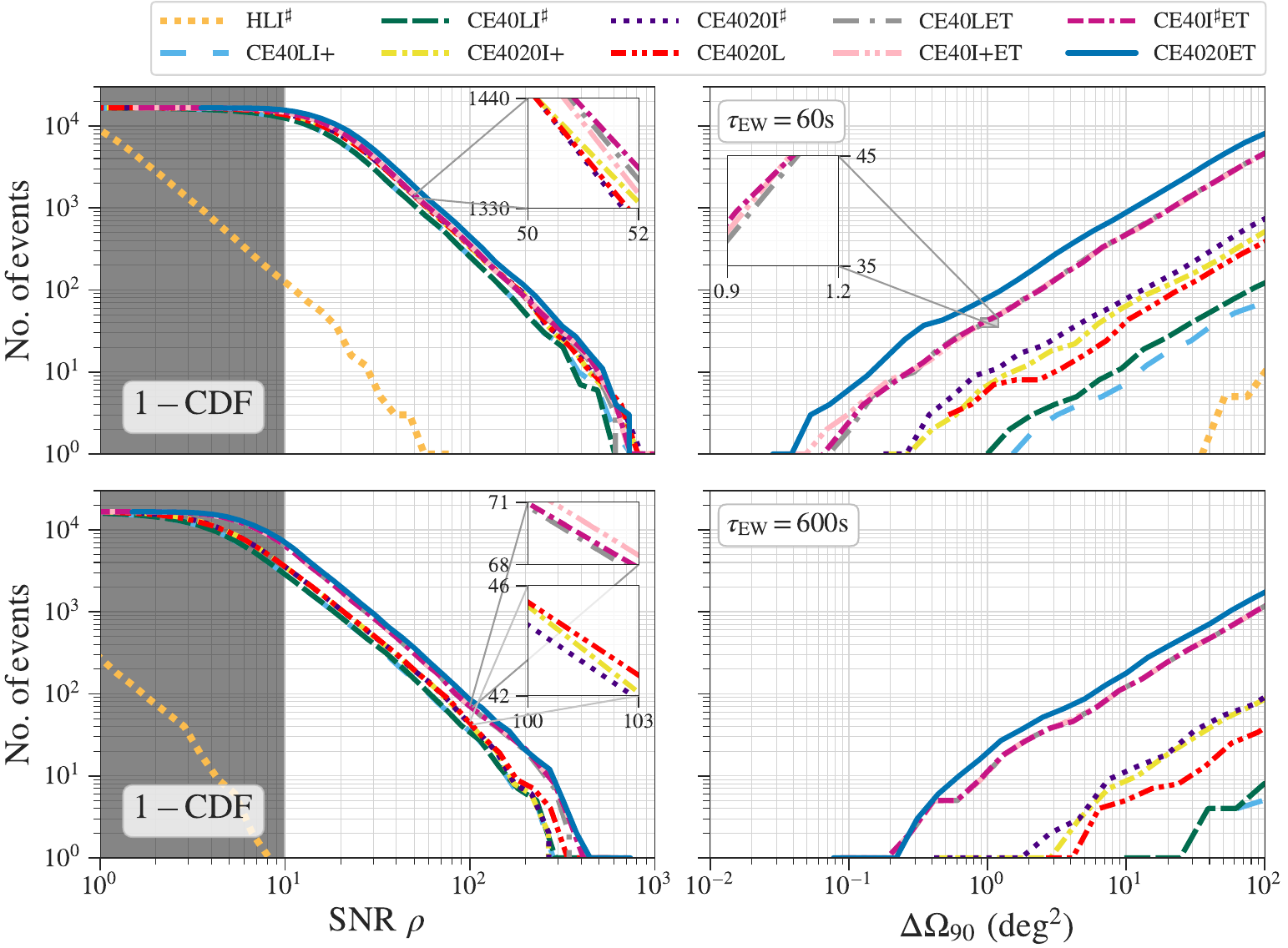}
  \caption{The number of events as a function of SNR (left panels) and sky localization area (right panels) for which early warning alerts can be issued at 1 minute (upper panels) and 10 minutes (bottom panels) before the BNS merger. Longer lead times with sufficient localization-precision allow telescope pre-positioning, improving the chances of detecting prompt GRB emission and early kilonova phases, while shorter windows require greater localization precision to enable these searches. Due to similar detection and localization capabilities, CE40I$^\sharp$ET, CE40I+ET, CE40LET are nearly indistinguishable. Similarly, CE4020I+, CE4020I$^\sharp$, CE4020L overlap in the SNR panels and they overlap with the aforementioned three in the top SNR panel. CE40LI+ and CE40LI$^\sharp$ also show nearly identical detection performance.} 
  \label{fig:ew_visible}
\end{figure*}

\begin{table*}[ht!]
\centering
\caption{Pre-Merger Sky Localization of BNS Coalescences Within 100, 10, and 1 $\mathrm{deg}^2$}
\vspace{-2mm}
\renewcommand{\arraystretch}{1.04}
\fontsize{9.6}{12}\selectfont
\begin{tabular}{c @{\hspace{3em}} ccc @{\hspace{3em}} ccc @{\hspace{3em}} ccc @{\hspace{3em}} ccc}
\toprule
EW Time & \multicolumn{3}{c}{$\tau_{EW} = 60\text{ s}$} & \multicolumn{3}{c}{$\tau_{EW} = 120\text{ s}$} & \multicolumn{3}{c}{$\tau_{EW} = 300\text{ s}$} & \multicolumn{3}{c}{$\tau_{EW} = 600\text{ s}$} \\
\hline
$\Delta \Omega \ (\text{deg}^2)$ & $\leq 100$ & $\leq 10$ & $\leq 1$ & $\leq 100$ & $\leq 10$ & $\leq 1$ & $\leq 100$ & $\leq 10$ & $\leq 1$ & $\leq 100$ & $\leq 10$ & $\leq 1$ \\
\hline
CE40LI$+$ & 51 & 9 & 3 & 29 & 6 & 2 & 12 & 4 & 3 & 2 & 0 & 0 \\
CE4020I$+$ & 400 & 53 & 9 & 258 & 35 & 7 & 137 & 21 & 4 & 68 & 10 & 3 \\
CE40I$+$ET & 3890 & 400 & 38 & 2881 & 295 & 27 & 1665 & 166 & 18 & 972 & 95 & 9 \\
\hline
HLI$^\sharp$ & 8 & 0 & 0 & 1 & 0 & 0 & 0 & 0 & 0 & 0 & 0 & 0 \\
CE40LI$^\sharp$ & 83 & 9 & 1 & 44 & 7 & 2 & 14 & 2 & 1 & 3 & 1 & 1 \\
CE4020I$^\sharp$ & 606 & 65 & 10 & 344 & 46 & 7 & 148 & 23 & 2 & 70 & 12 & 3 \\
CE40I$^\sharp$ET & 3965 & 413 & 38 & 2905 & 300 & 28 & 1669 & 168 & 18 & 973 & 95 & 9 \\
\hline
CE4020L & 297 & 34 & 4 & 182 & 20 & 8 & 77 & 9 & 2 & 34 & 7 & 3 \\
CE40LET & 3938 & 405 & 37 & 2898 & 297 & 27 & 1670 & 167 & 18 & 971 & 95 & 9 \\
CE4020ET & 7180 & 757 & 70 & 5261 & 548 & 53 & 2774 & 306 & 33 & 1480 & 149 & 15 \\
\hline
\end{tabular}
\label{tab:ew}
\end{table*}

Fig.~\ref{fig:ew_visible} shows the performance of various networks as a function of \ac{snr} and $\Delta \Omega_{90}$ at early warning times of $\tau_{ew} = 60\text{ s} \text{ and } 600\text{ s}$, and Table \ref{tab:ew} details the performance  at early warning times of $\tau_{ew} = 60\text{ s},\ 120\text{ s},\ 300\text{ s},\ 600\text{ s}$. The CE4020ET network consistently detects the highest number of events across all \ac{snr} values up to 10 minutes before merger, demonstrating its ability to observe both weaker signals and high-fidelity sources. It is followed by CE40I$^\sharp$ET, CE40LET, and CE40I$+$ET networks in performance. Other configurations also show strong detection capabilities, with the event count exceeding $10^4$ a minute before the merger and decreasing to a few thousand by 10 minutes pre-merger. In contrast, HLI$^\sharp$--lacking \ac{xg} detectors--detects $\sim100$ events at $\tau_{ew} = 60\text{ s}$ to near zero at $\tau_{ew} = 600\text{ s}$ at $\rho \geq 10$.

In terms of source localization, the number of detected events decreases by an order of magnitude from $\tau_{ew} = 60 \ \mathrm{s}$ to $\tau_{ew} = 600 \ \mathrm{s}$ across all detector configurations. CE4020ET detects the most precisely localized events, followed by CE40I$^\sharp$ET, CE40LET, and CE40I$+$ET. These networks detect several hundred events localized within $10 \ \mathrm{deg}^2$ at $\tau_{ew} = 60 \ \mathrm{s}$ and around 100 events at $\tau_{ew} = 600 \ \mathrm{s}$. Additionally, they detect about 10 events with $\Delta \Omega \leq 1 \ \mathrm{deg}^2$, 10 minutes before the merger, significantly improving the chances of capturing early transients by enabling precise electromagnetic follow-up. 
\begin{figure*}[htb!]
  \centering
  \includegraphics[width=.75\textwidth]{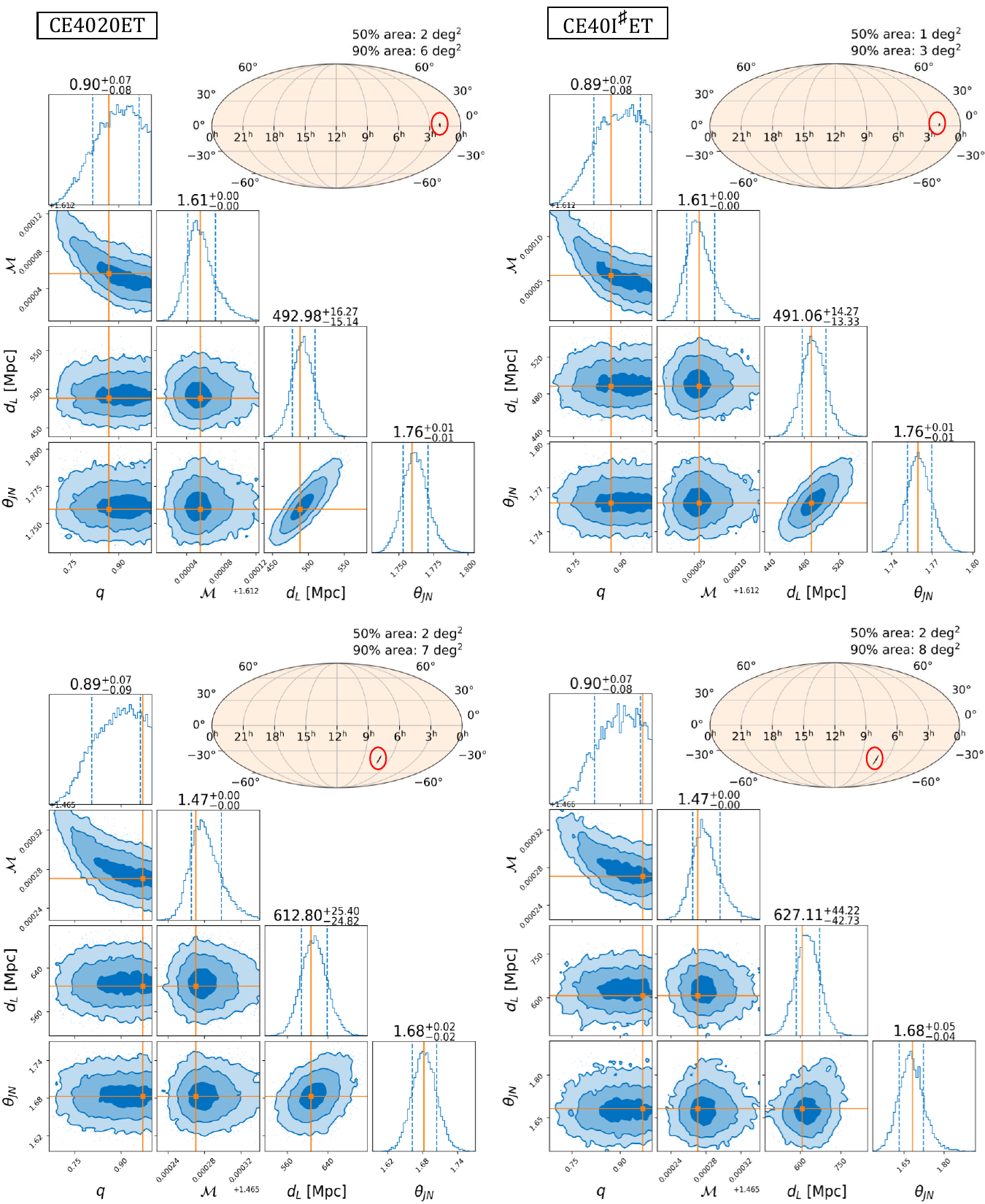}
  \caption{Parameter estimation corner plots and skymaps for two representative binary neutron star merger events in the CE4020ET (left) and  CE40I$^\sharp$ET (right) configurations. The corner plots show the posterior distributions for key source parameters, including mass ratio ($q$), chirp mass ($\mathcal{M}$), luminosity distance ($d_L$), and inclination angle ($\theta_{JN}$). The 1D marginalized distributions for each parameter are shown along the diagonal, while the off-diagonal plots show the 2D correlation between parameter pairs. The corresponding skymaps above each corner plot display the 50\% and 90\% credible regions for source localization in the sky.}
  \label{fig:corner_plots}
\end{figure*}

For early warnings, the CE4020L configuration performs similarly to CE4020I$+$, both detecting an order of magnitude fewer events than CE40I$^\sharp$ET within sky areas of $10$ and $100 \ \mathrm{deg}^2$ across all early warning times. The HLI$^\sharp$ network detects no events, while CE40LI localizes about a dozen events within $\Delta \Omega_{90} \leq 100 \ \mathrm{deg}^2$, 5 minutes before the merger.

These results underscore the substantial improvements in detection and localization achievable with \ac{xg} detectors and LIGO-India. While configurations including one or two \ac{ce} detectors with LLO and/or LIO provide useful early warning detections, networks such as CE4020ET, CE40IET, and CE40LET offer far greater scientific potential, enhancing opportunities for groundbreaking discoveries in multi-messenger astrophysics.

\begin{figure*}[htb!]
  \centering
  \includegraphics[width=0.78\textwidth]{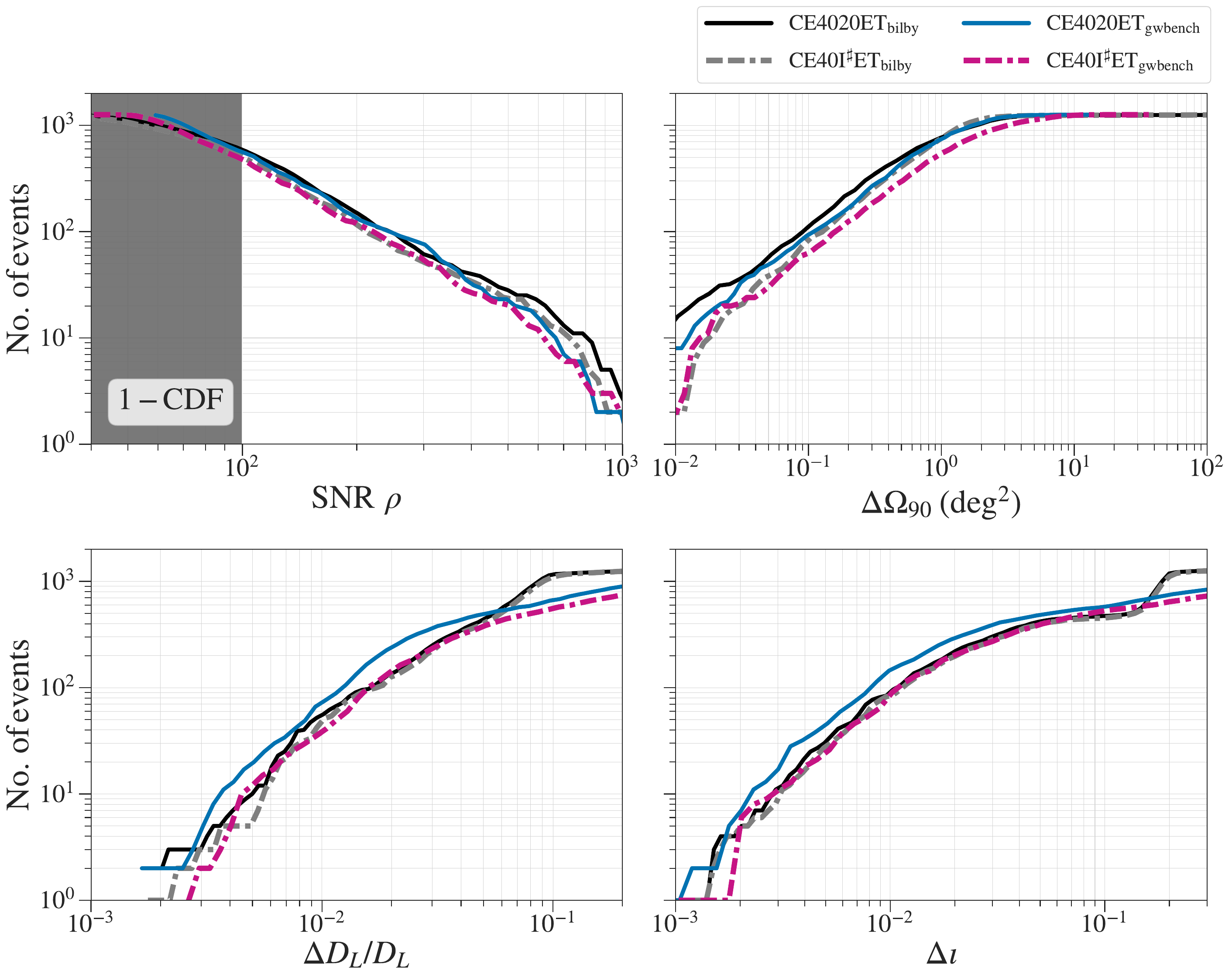}
  \caption{Comparative analysis of multi-messenger observational parameters for BNS mergers--representing $\sim$ 7.5\% of the annual BNS merger population--for optimal gravitational wave networks: CE4020ET and CE40I$^\sharp$ET. The figure presents BNS detection capabilities as a function of SNR, sky localization area, relative distance measurement uncertainty, and inclination angle measurement uncertainty. The black and dark grey lines represent the \BILBY analysis results for CE4020ET and CE40I$^\sharp$ET, respectively, while blue and pink lines correspond to the \textsc{\texttt{GWBENCH}} analysis for the same networks. The two networks exhibit comparable performance across all metrics, and the two analysis methods show close agreement–reinforcing confidence in the robustness of parameter recovery.}
  \label{fig:cdf_bilby}
\end{figure*}

\section{Verification of Optimal Networks' Performance Using \BILBY}
\label{sec:bilby}

To ensure the robustness and reliability of our conclusions, it is crucial to validate the results obtained using \texttt{\textsc{GWBENCH}} with alternative analysis techniques and pipelines. To that end, we conducted a cross-validation study using \BILBY--a Bayesian inference-based software package designed for gravitational wave data analysis--employing the relative binning technique for efficient likelihood computation and the PyMultiNest sampler for parameter estimation \citep{bilby_paper, pymultinest, relbin_bilby, relbin_cornish, relbin_zackay}. Focusing on the two optimal detector networks, CE4020ET and CE40I$^\sharp$ET, we selected a subset of $\sim 1250$ events--approximately 7.5\% of the total--from a yearly population of 16,585 gravitational wave events simulated by \texttt{GWBENCH}. This subset was selected from the 2000 highest \ac{snr} events in the CE4020ET network, of which 1249 were retained for analysis based on their injection values falling within the 90\% credible interval of the \BILBY posterior distributions for key source parameters. The geocentric time ($t_c$) in \BILBY was set to 1187008882.4 seconds, while all other injection parameters were kept consistent with the original \texttt{GWBENCH} simulations. We used the same data segment with an identical geocentric time for all \BILBY runs. This is not a concern as we analyze the simulated signals in zero-noise settings--free from stochastic noise fluctuations. Thus, any variation in parameter recovery stems solely from differences in the binary properties of the injected signals. As our simulated signal sources have varying sky locations, the detector response is different owing to the different relative orientations of the sources with the detector network despite the signals sharing the geocentric time. This controlled setup provides a consistent baseline for evaluating network performance without requiring multiple instantiations of GW data. 

Figure \ref{fig:corner_plots} highlights the comparative parameter estimation capabilities of the CE4020ET and CE40I$^\sharp$ET networks for two BNS merger events. The networks demonstrate similar posterior distributions and parameter correlations, indicating comparable performance in recovering key parameters such as chirp mass, mass ratio, and inclination angle. They also achieve similar source localization precision. For the first event, the 90\% credible sky areas ($\Delta \Omega_{90}$) are $6 \ \text{deg}^{2}$ for CE4020ET and $3 \ \text{deg}^{2}$ for CE40I$^\sharp$ET, with both networks achieving a relative uncertainty in luminosity distance of about 3\%. For the second event, $\Delta \Omega_{90}$ is $7 \ \text{deg}^{2}$ for CE4020ET and $8 \ \text{deg}^{2}$ with for CE40I$^\sharp$ET, while the relative uncertainties in luminosity distance are approximately 4\% and 7\%, respectively. This leads to some variation in three-dimensional localization volumes; however, the small values of both $\Delta \Omega_{90}$ and $\frac{\Delta D_L}{D_L}$ ensure limited practical impact. Consequently, both networks are highly effective for facilitating successful multi-messenger observations and host galaxy identification.

We also generated cumulative distribution function (CDF) and probability density function (PDF) plots for the verified events as shown in figures \ref{fig:cdf_bilby} and \ref{fig:pdf_bilby}, respectively. They offer a comparative assessment of the performance of CE4020ET and CE40I$^\sharp$ET networks using \texttt{GWBENCH} and \BILBY alalyses. The figures show that both networks exhibit similar distributions across key metrics, including \ac{snr}, sky localization area, relative distance measurement uncertainty, and inclination angle measurement uncertainty. These results confirm the highly comparable performance of CE4020ET and CE40I$^\sharp$ET in gravitational wave detection and parameter estimation. Furthermore, the close alignment between the results obtained using \BILBY and those from \texttt{GWBENCH} on this subset of events corroborates our earlier findings obtained for the annual BNS merger population with \texttt{GWBENCH} alone, thereby strengthening the overall credibility of our conclusions and affirming the suitability of CE4020ET and CE40I$^\sharp$ET as optimal networks for gravitational wave observations. Although this result may initially seem surprising, the fourfold increase in baseline between CE40 and LIGO-India--compared to the baseline of two CEs in the continental United States--accounts for the similar performance of the CE4020ET and CE40I$^\sharp$ET networks. Moreover, LIGO-India's location in a different plane relative to CE40 and CE20 improves the resolution of the two \ac{gw} polarizations, thereby breaking parameter degeneracies.

\vspace{2mm}
\section{Conclusions}

Our study highlights the pivotal role of LIGO-India within \ac{gw} detector networks consisting of \ac{xg} detectors, \ac{ce} and \ac{et}, emphasizing its importance to precision multi-messenger astronomy. Integrating LIGO-India alongside CE and ET substantially improves the detection capabilities, source localization precision, and early warning systems. Our findings show that, as expected, the CE4020ET configuration--comprising three \ac{xg} observatories--consistently performs best in all metrics evaluated. This configuration detects the highest number of events with the highest \ac{snr}, superior localization, and precise parameter estimation. The CE40I$^\sharp$ET configuration--which combines a 40-km CE, ET, and LIGO-India--achieves similar results due to its longer baseline and the improved polarization resolution provided by LIGO-India's unique geographic orientation.

In contrast, the CE4020L configuration, limited by its narrow baseline, exhibits inadequate sky localization capabilities, underscoring the necessity of including LIGO-India for optimal network performance. These findings have profound implications for astrophysical research. By enabling timely electromagnetic follow-ups of binary neutron star mergers owing to their superior detection and localization capabilities--even during late inspiral phases--optimal gravitational wave detector networks contribute in uncovering valuable insights into the nature of progenitors, the environment of host galaxies, and the behavior of matter under extreme conditions. Our results (Appendix \ref{app:aus}) also demonstrate that adding a Southern Hemisphere detector--either a 20 km CE or a 4km A$^\sharp$ interferometer in Australia--significantly enhances sky localization while maintaining parameter estimation performance, making an Australian site a compelling option for optimizing future \ac{gw} networks.

 While this work focuses on the impact of LIGO-India on multi-messenger astronomy, its inclusion within future networks will also be critical in addressing other scientific goals, such as testing general relativity and probing the nuclear equation of state. Nevertheless, our study demonstrates that LIGO-India will be indispensable in furthering our ability to study the universe and deepen our understanding of fundamental astrophysical phenomena.

\section{Acknowledgments}

The authors thank CE consortium members for their comments and valuable suggestions. The authors acknowledge the support through NSF grant numbers PHY-2207638, AST-2307147, PHY-2308886, and PHY-2309064. The authors acknowledge using the Gwave (PSU) and LDAS (CIT) clusters for computational/numerical work.

\bibliography{references}

\appendix
\section{Detection Efficiency}
\label{app:efficiency}

To analyze the detection capabilities of various \ac{gw} detector networks in observing \ac{bns} mergers, we estimate their detection efficiency and detection rate. The detection efficiency \(\epsilon(z, \rho^*)\) characterizes the detectable fraction of \ac{bns} mergers at redshift \(z\). It is calculated by distributing \ac{bns} signals across redshift intervals and counting the number of events exceeding a predefined \ac{snr} threshold \(\rho^*\). For each redshift interval, the efficiency is given by:

\begin{equation}
\epsilon(z, \rho^*) = \frac{1}{N_z} \sum_{k=1}^{N_z} H(\rho_k - \rho^*)
\end{equation}
where \(N_z\) represents the number of events, \(\rho_k\) is the \ac{snr} of the $k^{th}$ event, and $H(x)$ is the Heaviside step function. We use two \ac{snr} thresholds for our analysis: \(\rho^* = 10\) and \(\rho^* = 100\). \(\rho^* = 10\) is commonly adopted benchmark in \ac{gw} astronomy for a confident detection, whereas the higher threshold of \(\rho^* = 100\) represents a much stricter criterion focusing on the detection of rare, high-\ac{snr} events that will allow for precision science.

\begin{figure}[hbt]
  \centering
  \makebox[\textwidth]{%
        \includegraphics[width=.5\textwidth]{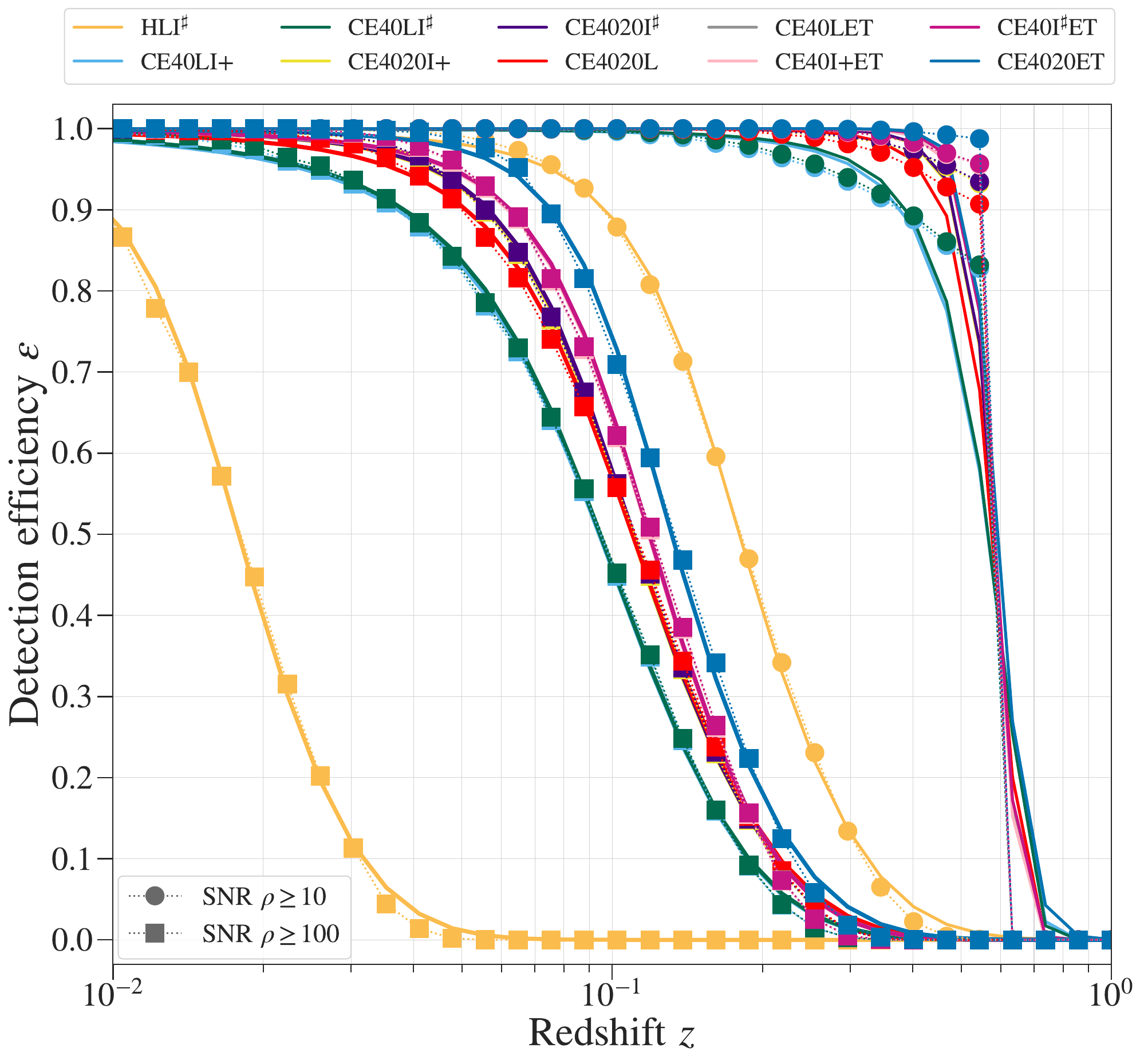}}
  \caption{The detection efficiency as a function of redshift for ten \ac{gw} detector networks. The solid circles represent values evaluated with a threshold SNR of 10, and the solid squares represent the values with a threshold SNR of 100. The solid lines are sigmoid fits for efficiency.}
  \label{fig:detection_eff}
\end{figure}

Figure \ref{fig:detection_eff} shows the efficiency of GW detector networks in observing \ac{bns} mergers as a function of $z$. At \(\rho^* = 100\),  most networks achieve a near-perfect efficiency at $z = 0.01$, with HLI$^\sharp$ attaining an efficiency of approximately 90\%. They observe almost all events with $\rho > 10$ up to $z = 0.2$, except for HLI$^\sharp$ whose efficiency drops below 50\% at that redshift. Notably, CE4020ET configuration detects almost the same number of events with $\rho \geq 100$ as HLI$^\sharp$ does with $\rho \geq 10$ at $z=0.05$. Moreover, networks with \ac{ce} and/or \ac{et} and LIGO-India maintain efficiencies of 70\% or more for events with $\rho \geq 10$ at $z = 0.5$. Remarkably, CE4020ET, CE40I$^\sharp$ET, CE40I+ET, and CE40LET observe $\sim 90 \%$ events under the same conditions. In contrast,  a network comprising only LIGO detectors detects only a few events at $z=0.5$ and is unlikely to detect any \ac{bns} mergers with $\rho \geq 100$ at $z = 0.07$. This suggests that all configurations are almost equally capable of detecting \ac{bns} mergers at near-Earth distances, but their efficiencies diverge with increasing redshift--especially under stricter \ac{snr} thresholds.

\section{Source Localization}

\begin{figure*}[htb!]
  \centering
  \includegraphics[width=0.9\textwidth]{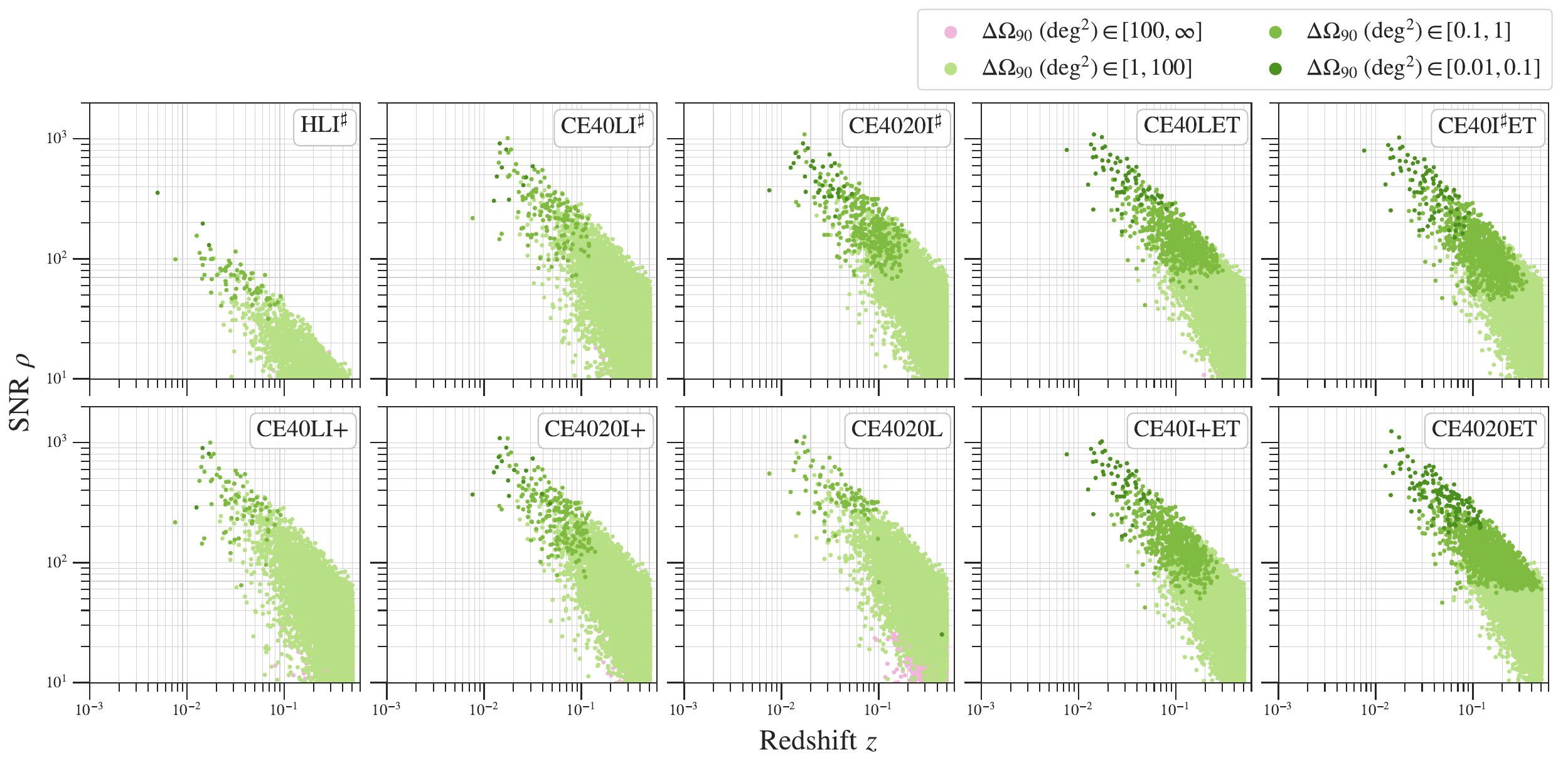}
  \caption{This figure depicts a comparison of the \ac{snr} ratio (SNR) for binary neutron star (BNS) mergers as a function of redshift (z) across various configurations of gravitational wave detectors. The colors represent different ranges for sky localization areas, indicating the precision with which the location of the gravitational wave sources can be pinpointed.}
  \label{fig:snr_z}
\end{figure*}

Figure \ref{fig:snr_z} shows the performance of the ten \ac{gw} detector networks by analyzing the relationship between \ac{snr}, $z$ and $\Delta \Omega_{90}$. The dark green points represent events with extremely precise source localization $ 0.01 \leq \Delta \Omega \leq 0.1 \deg^2$. The light green points correspond to $\Delta \Omega_{90} \in [1, 100] \deg^2$. Across the panels, the \ac{snr} expectantly decreases with increasing $z$. The combination of CE40 with other detectors (such as LIGO-India and \ac{et}) typically improves both the \ac{snr} and sky area. The spread and density of points in each panel highlight the variations in detector sensitivity and source localization precision under different operational circumstances and configurations.

\begin{table*}[ht]
\centering
\caption{Field of view (FOV) of selected electromagnetic telescopes}
\vspace{1mm}
\renewcommand{\arraystretch}{1.1}
\begin{tabular}{l c c }
\toprule
Telescope & FOV ($\deg^2$) & Reference \\
\hline
Rubin & 9.6 & \cite{Ivezi__2019}\\
EUCLID & 0.54 & \cite{Euclid_2022}\\
Athena & 0.35 & \cite{Athena_2013}\\
Roman & 0.28 & \cite{Chase_2022}\\
ngVLA (2.4 GHz; FWHM) & 0.17 & \cite{Murphy2018}\\
Chandra X-ray Observatory & 0.15 & \cite{Weisskopf:2000tx}\\
Lynx & 0.13 & \cite{Gaskin2019}\\
Swift–XRT & 0.12 & \cite{Burrows2000}\\
Keck Observatory & 0.11 & \cite{Bundy_2019}\\
Giant Magellan Telescope (GMT) & 0.11 & \cite{Johns_2012}\\
Extremely Large Telescope (ELT)$^a$ & 0.03 & \cite{Padovani:2023dxc}\\
Jansky VLA (3 GHz; FWHM) & 0.0625 & \cite{Perley_2011}\\
\hline
\end{tabular}
\label{tab:telescopes}
\begin{flushleft}
\justifying\textbf{Note.} $^a$ Field of view of the Multi-AO Imaging Camera for Deep Observations (MICADO) instrument.
\end{flushleft}
\end{table*}

\begin{figure*}[hbt!]
  \centering
  \includegraphics[width=0.85\textwidth]{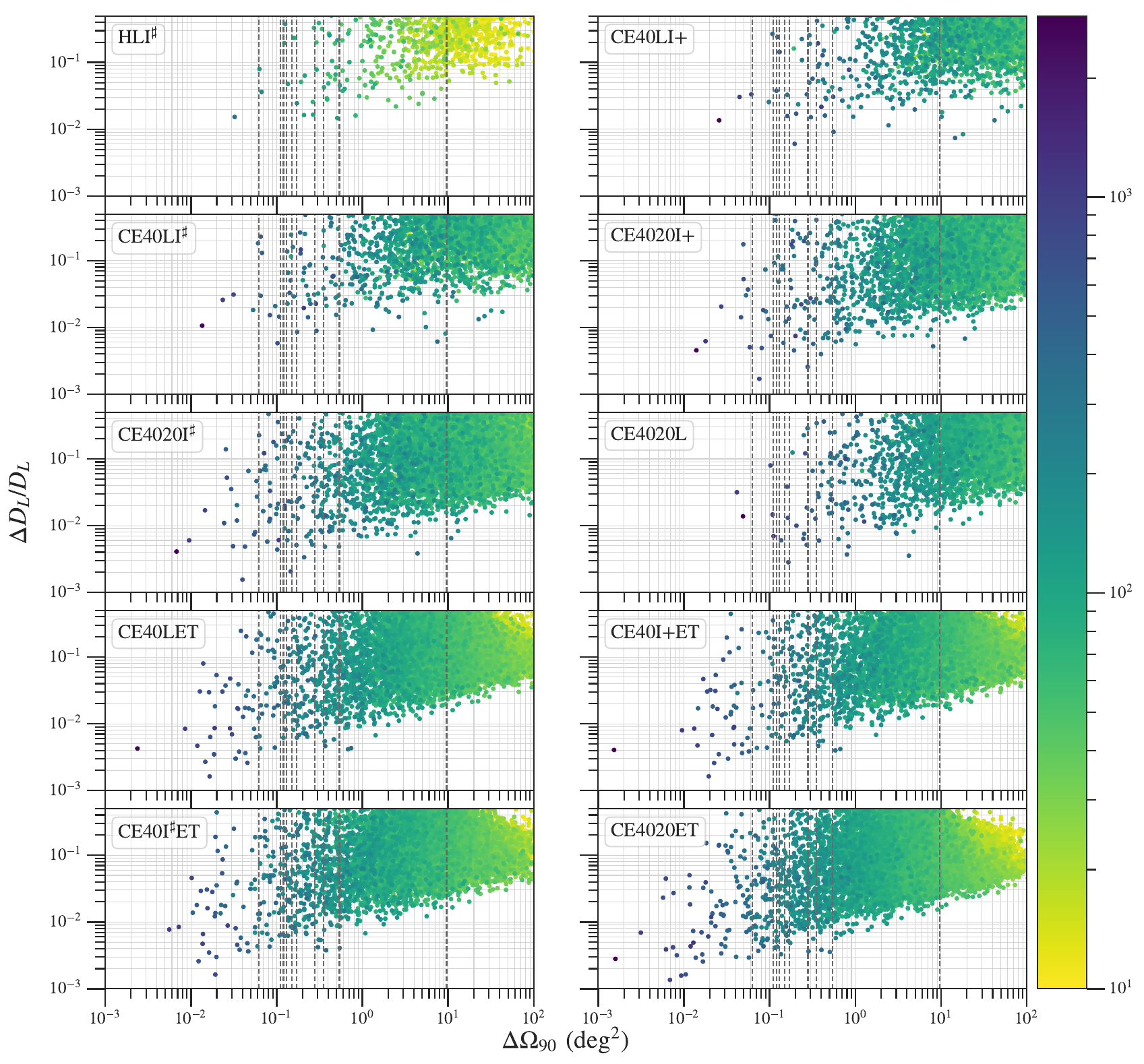}
  \caption{This figure depicts the relative uncertainty in luminosity distance as a function of 90\% credible sky area in square degrees for BNS mergers for different gravitational wave detector configurations. The color denotes the signal's network \ac{snr}.}
  \label{fig:dist_area}
\end{figure*}

Figure \ref{fig:dist_area} presents the 3D localization capabilities of our networks, with the color gradients representing SNR $\rho$ and the dotted lines representing the field of view (FOV) of selected electromagnetic telescopes listed in Table \ref{tab:telescopes}. The CE4020ET and CE40I$^\sharp$ET networks achieve precise 2D localization areas and smaller errors in $D_L$, with majority of events localized within $10 \ \text{deg}^2$ and achieving $\Delta D_L/D_L \leq .1$. This significantly enhances the efficiency of follow-up observations. HLI$^\sharp$, CE40LI, and CE4020L networks, while still capable of localizing events within $100 \ \text{deg}^2$, exhibit higher luminosity distance errors and detect fewer events with a more precise localization. Hence, they are less compatible with follow-up observations using instruments with a narrow FoV. 

\section{PDF Distributions}

\begin{figure}[h]
  \centering
  \includegraphics[width=0.80\textwidth]{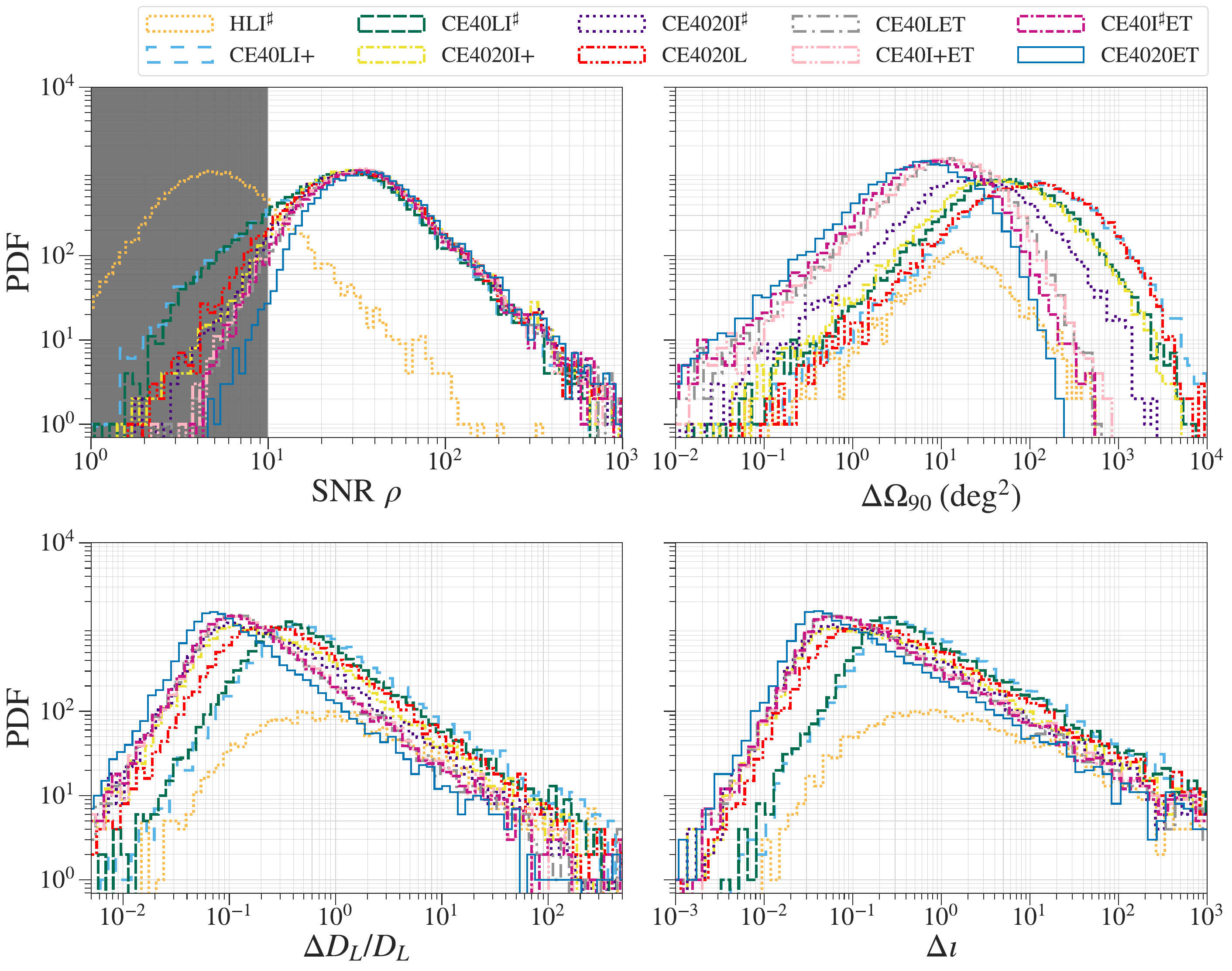}
  \caption{Comparative analysis of multi-messenger observational parameters for BNS mergers across ten gravitational wave detector network configurations. The figure presents multiple panels, each illustrating the Probability Density Function (PDF) distributions of BNS detection capabilities as function of SNR, sky localization area, relative distance measurement uncertainty, and inclination angle measurement uncertainty, respectively.}
  \label{fig:visible_mma_pdf}
\end{figure}

Figure \ref{fig:visible_mma_pdf} presents the probability density functions (PDFs) for key performance metrics across different gravitational-wave detector network configurations, complementing the cumulative distribution functions (CDFs) shown in Fig. \ref{fig:visible_mma}. These PDFs offer a detailed characterization of the event distributions with respect to signal-to-noise ratio (SNR) $\rho$, 90\% credible sky area $\Delta \Omega_{90}$, relative luminosity distance uncertainty $\Delta D_L / D_L$, and inclination angle measurement uncertainty $\Delta \iota$. The networks CE4020ET and CE40I$^\sharp$ET demonstrate superior signal detection capabilities and source localization accuracies. Their SNR distributions peak in the tens, and their PDFs for $\Delta \Omega_{90}$ peak at a few square degrees, indicating precise sky localization. These networks also achieve high precision in measuring luminosity distance and inclination angle, as evidenced by PDFs skewed toward smaller uncertainties.

In contrast, the HLI$^\sharp$ configuration exhibits limited sensitivity, with most events observed at SNRs below the detection threshold of 10. It also shows higher errors in luminosity distance and inclination angle measurements, and broader localization areas. The CE4020L network shows improvement over HLI$^\sharp$ but still underperforms compared to the CE4020ET and CE40I$^\sharp$ET configurations. These results underscore that CE4020ET and CE40I$^\sharp$ET are the optimal network configurations, highlighting the critical importance of including LIGO-India in the detector network.

\begin{figure}
  \centering
  \includegraphics[width=0.80\textwidth]{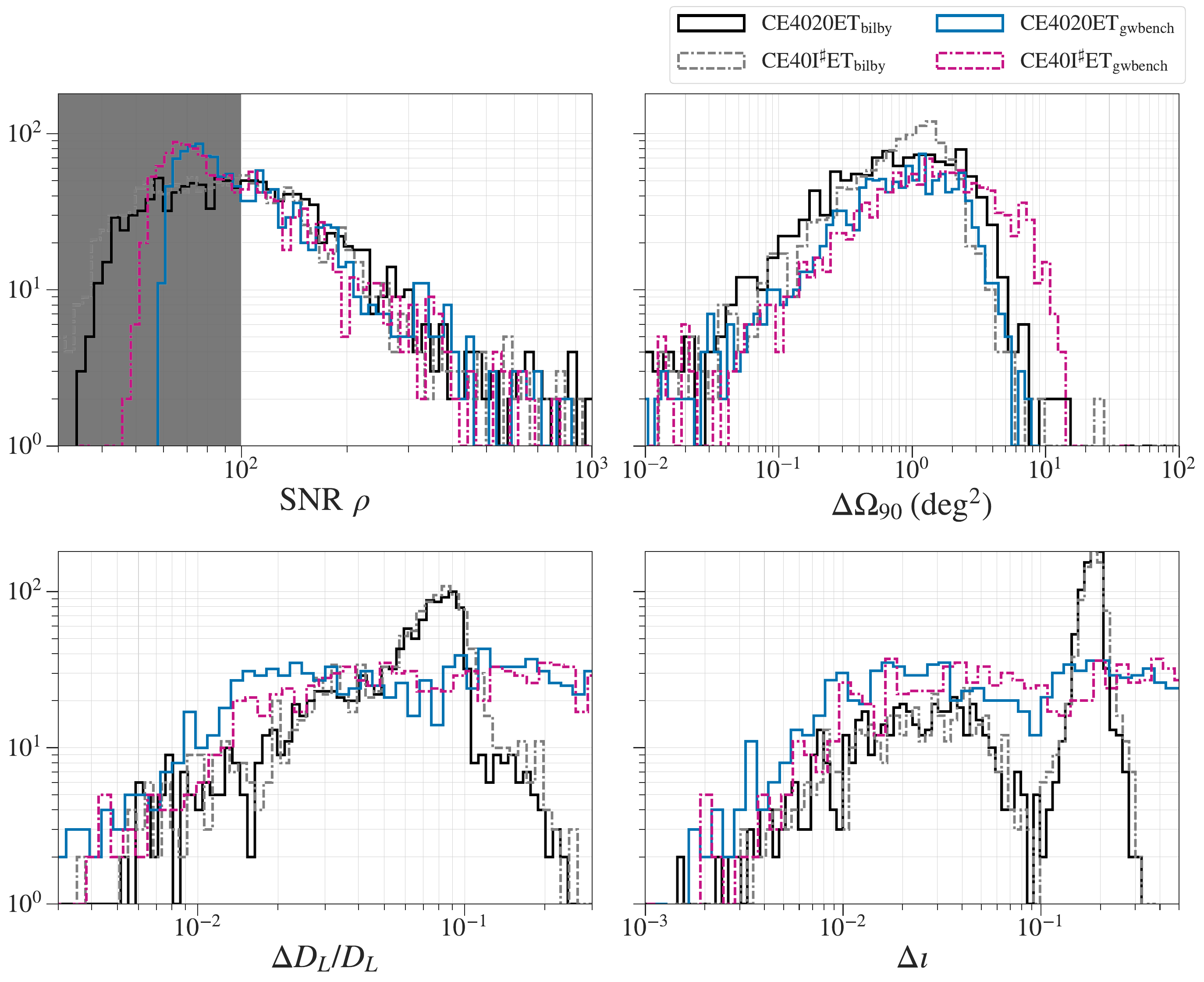}
  \caption{Probability density function (PDF) distributions for key observational parameters of BNS mergers, representing $\sim$ 7.5\% of the annual BNS merger population, as detected by CE4020ET and CE40I$^\sharp$ET networks. The panels present BNS detection capabilities as a function of SNR, sky localization area, relative distance measurement uncertainty, and inclination angle measurement uncertainty, respectively. Results are shown for two methods: \texttt{GWBENCH} (blue for CE4020ET and pink for CE40I$^\sharp$ET) and \BILBY (black and dark grey for the respective networks).}
  \label{fig:pdf_bilby}
\end{figure}

To further assess the performance of these optimal networks, Figure \ref{fig:pdf_bilby} compares the results obtained using \texttt{gwbench} and \texttt{Bilby} for a subset of events. Both methods exhibit strong consistency in SNR and sky localization area. Some differences appear in luminosity distance and inclination angle measurements; however, both analyses yield small errors in these metrics for CE4020ET and CE40I$^\sharp$ET. While we did not perform an event-by-event comparison for the two methods, the broader population-level agreement captured in the CDF and PDF plots across these key metrics underscores the reliability of both approaches. This consistency further supports the importance of including LIGO-India in the network configuration.

\section{Network Variants with a Southern-Hemisphere Observatory}
\label{app:aus}

In this appendix, we evaluate the potential benefits of adding a gravitational-wave detector in the Southern Hemisphere to enhance the overall performance of the global detector network. Australia has long been recognized as a favorable location for such an expansion, with proposed sites near the Gingin facility and in New South Wales~\citep{Searle_2006, Blair_2008, Wen_2010, Evans_2021_CosmicExplorer, Srivastava_2022, Gardner_2023}. We consider four alternative network configurations that include either a 20 km Cosmic Explorer–class interferometer (denoted S20) or a 4 km detector operating at A$^\sharp$ sensitivity (denoted S$^\sharp$), located at approximately 146° East, 32° South. The definitions of these additional network configurations are summarized in Table~\ref{tab:alternate_configurations}.

\begin{table}[htb!]
  \centering
  \caption{Alternate Gravitational Wave Detector Networks with a Southern-Hemisphere Detector in Australia}
  \vspace{1.5mm}
  \renewcommand{\arraystretch}{1.04}
  \begin{tabular}{c c}
    \hline\hline
    Configuration & Detectors Included \\
    \hline 
    CE4020S$^\sharp$ & 40 km CE + 20 km CE + S in A$^\sharp$ \\
    CE40S$^\sharp$ET & 40 km CE + S in A$^\sharp$ + ET \\
    CE40S20I$^\sharp$ & 40 km CE + 20 km CES + LIO in A$^\sharp$  \\
    CE40S20ET & 40 km CE + 20 km CES + ET \\
    \hline
  \end{tabular}
  \begin{flushleft}
\centering\textbf{Notes.} S20 refers to a 20 km CE-class detector in Australia and S refers to a 4 km detector in Australia. 
 \end{flushleft}
 \label{tab:alternate_configurations}
\end{table}

\begin{figure}
  \centering
  \includegraphics[width=0.80\textwidth]{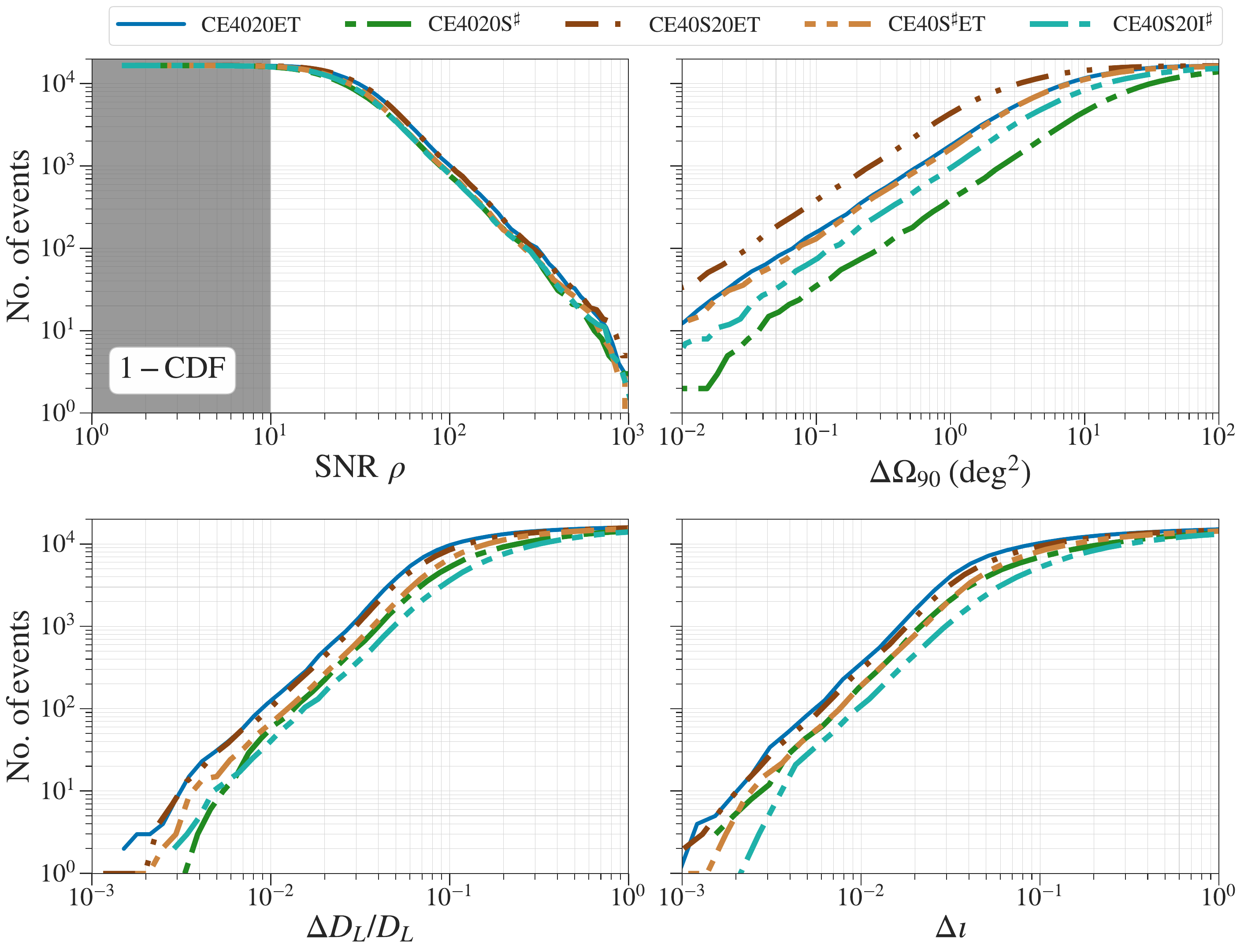}
  \caption{Cumulative distribution functions of SNR, 90\% credible localization area, and parameter uncertainties for networks incorporating a southern‑hemisphere detector in Australia (S20 and S$^\sharp$ variants), with CE4020ET shown for reference.}
  \label{fig:aus_cdfs}
\end{figure}

\begin{figure}
  \centering
  \includegraphics[width=1.\textwidth]{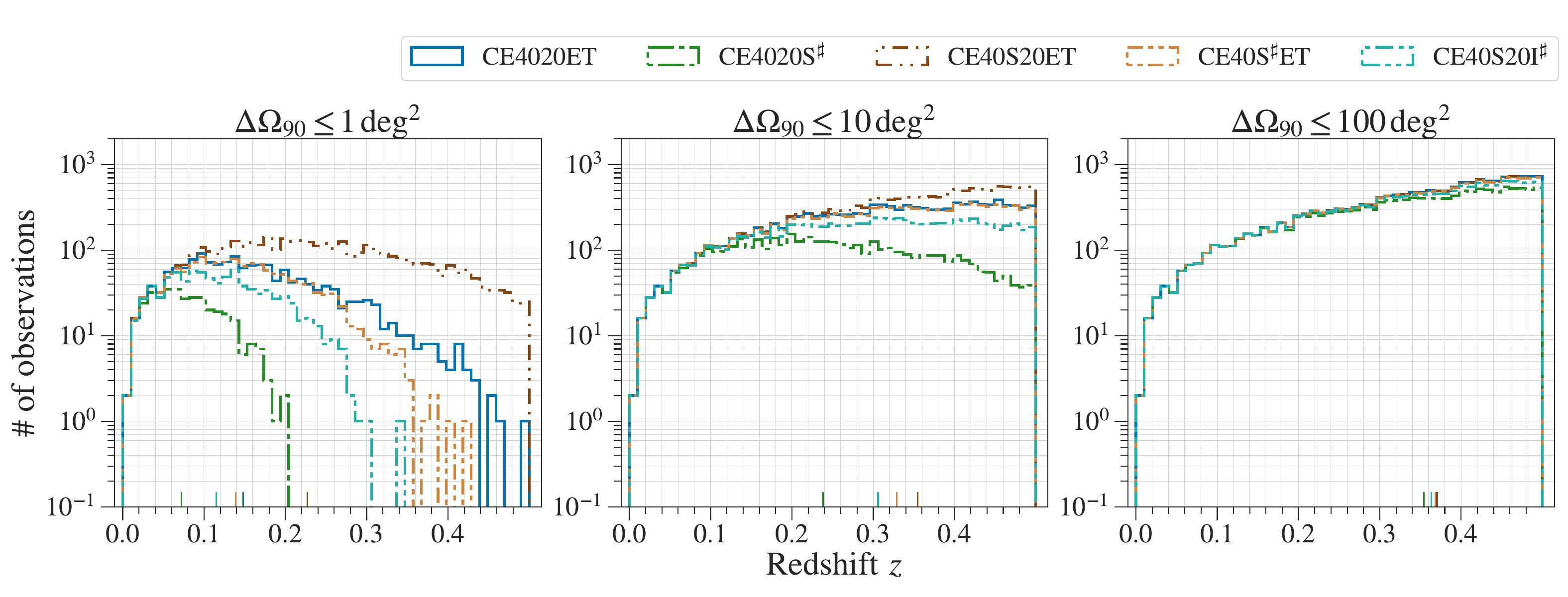}
  \caption{Number of events that can be localized within sky areas of 1 (left), 10 (middle), and 100 square degrees (right), respectively, as a function of redshift for four networks incorporating a southern-hemisphere detector in Australia. The CE4020ET network is shown for reference.}
  \label{fig:aus_localization}
\end{figure}

Performance is quantified using (i) cumulative distribution functions of SNR, 90\% credible sky‑localization area, and uncertainties on luminosity distance and inclination angle (Fig.\ref{fig:aus_cdfs}), and (ii) histograms of the number of events localized within 1, 10, and 100 deg$^2$ as a function of redshift (Fig.\ref{fig:aus_localization}). For reference, curves for the baseline CE4020ET network are included. The SNR panel in \ref{fig:aus_cdfs} shows that replacing the North American CE20 in CE4020ET with the one in Australia (S20) leaves the SNR distribution essentially unchanged. Other networks: CE4020S$^\sharp$, CE40S$^\sharp$ET and CE40S20I$^\sharp$ deliver marginally lower detection performance. 

The top right panel illustrates that CE40S20ET localizes approximately 30 events within $0.01 \deg^{2}$ and $\sim$3900 events within $1 \deg^{2}$--more than double the counts for CE4020ET. CE40S$^\sharp$ET shows identical performance to CE4020ET, whereas CE40S20I$^\sharp$ localizes roughly half as many events within areas $\leq 1 \deg^{2}$ and CE4020S$^\sharp$ underperforms by a factor of 4 or more. The redshift-dependent histograms in Fig.~\ref{fig:aus_localization} show that all network configurations perform similarly when using a localization threshold of $\leq$100 deg$^2$. However, significant differences emerge under more stringent thresholds. At $z = 0.2$, the CE40S20ET network localizes over 100 events within 1 deg$^2$, compared to approximately 50 events for both CE4020ET and CE40S$^\sharp$ET, around 30 for CE40S20I$^\sharp$, and only a handful for CE4020S$^\sharp$.  All networks but CE4020S$^\sharp$ exceed 100 detections within $\Delta\Omega_{90} \leq10 \deg^{2}$ at $z=0.5$, with CE40S20ET localizing $\sim500$, CE4020ET and CE40S$^\sharp$ET $\sim300$, CE40S20I$^\sharp$ $\sim200$, and CE4020S$^\sharp$ localizing only $\sim40$ events. Finally, the lower panels in Fig. \ref{fig:aus_cdfs} show that CE40S20ET recovers both luminosity distance and inclination angle with similar uncertainties as CE4020ET, with CE40S$^\sharp$ET and CE4020S$^\sharp$ closely trailing and CE40S20I$^\sharp$ recovering $\sim$30\% of events with those uncertainties.

These results confirm that a southern‑hemisphere facility would powerfully complement existing northern detectors. In particular, deploying a 20 km CE interferometer in Australia (S20) doubles the number of events localized within stringent sky‑area thresholds while preserving the CE4020ET network’s parameter‑recovery performance. Even adding a 4 km Australian detector operating at A$^\sharp$‑sensitivity in a network with CE40 and ET yields localization comparable to CE4020ET with only marginal impact on distance and inclination uncertainties. Therefore, a location in Australia should be considered to fully realize the localization and inference potential of \ac{gw} networks.

\end{document}